\documentclass[pre, aps, showpacs]{revtex4}

\usepackage{epsfig}
\usepackage{amstext}
\usepackage{amsbsy}
\usepackage{amsmath}
\usepackage{xcolor}

\begin{document}

\title{Large Pseudo-Counts and $L_2$-Norm Penalties Are Necessary\\ for the Mean-Field Inference of Ising and Potts Models}

\author{J.~P.~Barton$^{1,2}$, S.~Cocco$^3$, E.~De Leonardis$^{3,4}$, R.~Monasson$^5$}
\affiliation{$^1$ Department of Chemical Engineering, MIT, Cambridge, Massachusetts 02139, USA\\
$^2$ Ragon Institute of MGH, MIT and Harvard, Boston, Massachusetts 02129, USA\\
$^3$ Laboratory of Statistical Physics of the Ecole Normale Sup\'erieure, associated to CNRS and University P\&M. Curie, 24 rue Lhomond, 75005 Paris, France\\
$^4$ UMR 7238, Computational and Quantitative Biology,  UPMC Univ Paris 06, France Sorbonne Universit\'es, 75005 Paris, France \\
$^5$ Laboratory of Theoretical Physics of the Ecole Normale Sup\'erieure, associated to CNRS and University P\&M. Curie, 24 rue Lhomond, 75005 Paris, France}
\date{\today}

\begin{abstract}
Mean field (MF) approximation offers a simple, fast way to infer direct interactions between elements in a network of correlated variables, a common, computationally challenging problem with practical applications in fields ranging from physics and biology to the social sciences. However, MF methods achieve their best performance with strong regularization, well beyond Bayesian expectations, an empirical fact that is poorly understood. In this work, we study the influence of pseudo-count and $L_2$-norm regularization schemes on the quality of inferred Ising or Potts interaction networks from correlation data within the MF approximation. We argue, based on the analysis of small systems, that the optimal value of the regularization strength remains finite even if the sampling noise tends to zero, in order to correct for systematic biases introduced by the MF approximation. Our claim is corroborated by extensive numerical studies of diverse model systems and by the analytical study of the $m$-component spin model, for large but finite $m$. Additionally we find that pseudo-count regularization is robust against sampling noise, and often outperforms $L_2$-norm regularization, particularly when the underlying network of interactions is strongly heterogeneous. Much better performances are generally obtained for the Ising model than for the Potts model, for which only couplings incoming onto medium-frequency symbols are reliably inferred. 
\end{abstract}


\maketitle

\section{Introduction}

Inferring effective interaction networks from the measured time-series of a population of variables is a problem of increasing importance across multiple scientific disciplines, including biology (for the study of protein residue coevolution \citep{Socolich:2005js, Weigt:2009ba, Burger:2010, Morcos:2011ct, Hopf:2012kv, Cocco:2013pl, Ferguson:2013kb}, gene networks \citep{Segal:2003vq, Lezon:2006, Bonneau:2007kj}, neuroscience \citep{Schneidman:2006vz, Field:2007is, Pillow:2008bo, Cocco:2009hw, Tkacik:2010}, and ecology \citep{Faust:2012kq, Friedman:2012vs, Bialek:2012, Hekstra:2013}, among others), sociology \citep{Christakis:2007fy, Lazer:2009ci, Borgatti:2009vk}, and finance \citep{Plerou:1999wh, Moro:2009fh}. One popular approach to this problem is to infer a simple graphical model which reproduces the low-order stationary statistics of the data, such as the single variable frequencies and the pairwise correlations. Inferred model parameters then give clues about the underlying network of interactions between the variables. 

When the 'true' model is uncertain, in practice one often searches for the maximum entropy, or least constrained \citep{Jaynes:1982wh}, model capable of reproducing the data. The Ising model is the maximum entropy model capable of reproducing the one- and two-point statistical constraints between binary variables, \textit{e.g.}~the activity of a population of neurons, which are either silent or emit a spike within a certain time window. When the variables take more than two values, for example specifying the amino acid present at each site in a protein sequence, the Potts model is a natural extension. In both cases the computational problem consists of finding the set of couplings $J_{ij}(a,b)$, expressing the interactions between the 'symbol' $a$ of variable $i$ and the symbol $b$ of variable $j$, from the measured correlations $c_{ij}(a,b)$. This is referred to as the inverse Ising, or Potts, problem. An exact solution generally requires computational efforts increasing exponentially with the system size (number of spin variables) $N$. Efficient and accurate approximation schemes are thus required for the analysis of real data, and host of methods have been developed with this goal in mind \citep{Hastie:2009wp, Monasson:2009ts, Nguyen:2011kk, SohlDickstein:2011im, Cocco:2011fo, Ravikumar:2010wa, Aurell:2012we}. 

Among the algorithms developed for solving the inverse Ising and Potts problems, the mean-field (MF) inference procedure is certainly the simplest. MF is fast as it runs in a time growing polynomially with $N$. MF simply approximates the coupling matrix with minus the inverse of the correlation matrix, a result which would be exact for Gaussian distributed variables, but which is only approximate for the Ising or the Potts model. This method has been shown to give very good results for various biologically-motivated problems, such as the study of amino acid covariation in proteins \citep{Morcos:2011ct, Hopf:2012kv, Cocco:2013pl} and multi-electrode recordings of neural activity \citep{Barton:2013fj}. 

Despite its popularity, key ingredients for the success of MF inference remain poorly understood. In particular an essential ingredient of the inference from real data is the presence of a regularization term ensuring that the  inverse problem be always well defined. To this aim $L_1$- or $L_2$-norm regularization of the couplings, or pseudo-count regularization of the correlations can be used \cite{Hastie:2009wp}. However, from a Bayesian point of view the optimal strength of these regularization terms is expected to decrease with the level of the noise, and should vanish in the limit of perfect sampling.  Empirical studies show that this is not the case  for MF inference: the best performance of MF inference is achieved only in the presence of very strong regularization. Predictions of contacts between residues on protein folds based on MF inference are optimal when the regularization terms are very strong, without any apparent dependence on the number of data \citep{Morcos:2011ct, Marks:2011hj, Hopf:2012kv, Cocco:2013pl}. Another related finding which has lacked any explanation so far is why pseudo-counts are generally better than other regularization schemes, such as $L_2$- or $L_1$-norm regularization of couplings, when combined with MF inference.

Here we explore the performance of various regularization schemes for MF inference on Ising and Potts model, and the reasons behind their success, through the analytical analysis of small model spin systems combined with extensive numerical studies of larger systems. First, we show that abnormally strong regularization is necessary to correct for errors produced by the MF approximation itself, which we explain using the simplest case of models with few variables ($N=2,3$). In addition we show that, in systems with homogeneous interactions, pseudo-counts and $L_2$-norm regularization are performing similar functions, but not $L_1$. 
When the (Ising or Potts) model includes a large number $N$ of spin variables we show based on numerical simulations that the same phenomenon takes place: large regularization is necessary, but pseudo-counts do a better job than $L_2$ for strongly heterogeneous networks. We explain why this is so using analytical arguments, based on the analysis of the $O(m)$ continuous spin model for large but finite $m$. MF is exact for this model in the $m\to\infty$ limit, and we show that the optimal pseudo-count remains finite in the absence of sampling noise: the optimal penalty is of the order of $\frac{1}{m}$, which estimates the deviation of the model with respect to Gaussianity.  Moreover
 inference is less affected by sampling noise when using large pseudo-count than when using large $L_2$-norm.
Finally we show that inference performances, even with large pseudo-count,   may be much poorer for the Potts than for the Ising model, especially so when the symbols on each site largely differ in their frequencies.   Our study therefore provides a strong basis for the use of large regularization penalties with mean-field inference, which thus far had been totally empirical.

The paper is organized as follows. In Section II we present the different regularization schemes studied in the paper, and briefly recall how couplings are inferred from correlations within the Gaussian (MF) approximation. In Section III we present a detailed analysis of the error of inference due to MF and how those errors are corrected, with varying success, in the presence of regularization. Section IV reports the performances of the MF as a function of the regularization strength based on extensive numerical simulations of Ising and Potts models and with diverse interaction distributions and structures. The statistical mechanics of the $O(m)$ model and results regarding the optimal value of the regularization penalty with MF inference are presented in Section V. Conclusions are proposed in Section VI.

\section{Reminder on mean-field inference and regularization}

The mean-field approximation consists, as far as inference is concerned, of approximating the Ising (or the Potts) model couplings with the off-diagonal elements of minus the inverse of the correlation matrix. We recall below that this result can be found for the Ising model within the Gaussian approximation (Section~\ref{approxgauss}), where the discrete nature of the spin variables is omitted. Section~\ref{schemes} briefly presents the regularization schemes studied here, namely the pseudo-count and the $L_2$-norm (as well as the $L_1$-norm). Specificities of the inference applied to the Potts model are discussed in Section \ref{potts934}.

\subsection{Inference of couplings within the Gaussian approximation} \label{approxgauss}

In the Gaussian approximation the differences between the Ising spin variables $\sigma_i=0,1$ and their empirical average values,  
\begin{equation}
\hat\sigma_i=\sigma_i - \langle \sigma_i\rangle \ ,
\end{equation}
are assumed to be drawn from a Gaussian distribution, with zero mean and empirical covariance matrix $c_{ij}=\langle \hat \sigma _i \hat \sigma_j \rangle=\langle \sigma _i \sigma_j \rangle-\langle \sigma _i \rangle\langle \sigma_j \rangle$. The likelihood of a configuration is 
\begin{equation}\label{measure}
P(\hat\sigma_1,\ldots ,\hat\sigma_N)= \frac{ \sqrt{\det  \hat J}}{ ( 2\, \pi) ^{N/2}}\; \exp \left( -\frac 12 \sum_{i,j} \hat J_{ij} \, \hat \sigma_i \, \hat \sigma_j \right)\ ,
\end{equation}
where the off-diagonal elements of the $\hat J$ matrix coincide with the opposite of the couplings: $\hat J_{ij}=-J_{ij}$, for all $i\ne j$. Contrary to the Ising or Potts model the diagonal elements of $\hat J$ are important to define the measure (\ref{measure}), and their values will be specified later on.

In the formulas above $\langle\cdot\rangle$ denotes the empirical average over the data set, composed of $B$ independently sampled configurations of the model.  The log-likelihood of the data within the Gaussian model (\ref{measure}) is a function of its empirical covariance, given by
\begin{equation}\label{llmf}
L(c|J) =  \frac B2\,\big( -\text{trace}( \hat J\, c)+ \log \det \hat J  \big) 
\end{equation}
Maximization of $L$ over $\hat J$ for a fixed $c$ gives $\hat J=c^{-1}$. For the off-diagonal entries we obtain
\begin{equation}\label{inf}
J_{ij} = -\hat J_{ij} = - (c^{-1})_{ij}\ ,
\end{equation}
while the diagonal couplings $\hat J_{ii}$ are Lagrange parameters enforcing the $N$ conditions $\langle \hat\sigma_i^2\rangle=\langle \sigma_i\rangle(1-\langle \sigma_i\rangle)$ for $0,1$ spins, or $\langle \hat\sigma_i^2\rangle=1-\langle \sigma_i\rangle^2$ for $\pm 1$ spins. Hence, inference with the Gaussian model provides the same expression for the couplings than the mean-field approximation. In the following we will use the subscript MF  to refer to the couplings given by expression (\ref{inf}).
 
\subsection{Regularization schemes} \label{schemes}

Empirical averages are computed from a finite number $B$ of configurations, and therefore the correlation matrix $c$ is not always invertible. Zero modes are found when some configurations of variables are never sampled, {\em e.g.}~when variables $\sigma_i$ and $\sigma_j$ are equal in all $B$ configurations. The invertibility of a the correlation matrix can be ensured by introducing some form of regularization to the model. From a Bayesian point of view, such terms can be thought of as the contribution of prior distributions for the model parameters, and their amplitude should vanish in the limit of perfect sampling ($B\to\infty$). Below we review three popular regularization schemes: pseudo-count, and the $L_1$- and $L_2$-norm regularization of couplings. For the sake of simplicity, definitions are given for the Ising model, the extension to the Potts case being straightforward.

\subsubsection{Pseudo-count}
A very simple regularization scheme consists in adding extra `pseudo' observations to the real data in order to cure singularities caused by strong correlations. For instance, if $\sigma_i=0$ in all $B$ configurations in the real data, $\sigma_i$ could be given value 1 in a pseudo $(B+1)^{th}$ configuration. This is the so-called `pseudo-count' method, popular in the analysis of protein sequence data in biology, which can be interpreted in terms of a Dirichlet prior distribution for the observation of each valuethe $\sigma_i$ \citep{durbin1998biological}. Typically, one chooses a prior distribution in which each value of $\sigma_i$ is considered equally likely. In this case the pseudo-count changes the frequencies and correlations in the following way:
\begin{equation}
\langle \sigma_i \rangle \to (1-\alpha)\, \langle \sigma_i\rangle + \frac {\alpha}2\ , \quad \langle \sigma_i\sigma_j\rangle \to (1-\alpha)\, \langle \sigma_i\sigma_j\rangle + \frac {\alpha}4 \quad (i\ne j)\ , 
\end{equation}
for $0, 1$ spins, and 
\begin{equation}\label{defpc2}
\langle \sigma_i \rangle \to (1-\alpha)\, \langle \sigma_i\rangle \ , \quad \langle \sigma_i\sigma_j\rangle \to (1-\alpha)\, \langle \sigma_i\sigma_j\rangle \quad (i\ne j)\ , 
\end{equation}
for $\pm 1$ spins. Diagonal terms are constrained to $\langle \sigma_i ^2\rangle =\langle \sigma_i\rangle$ for $\sigma_i=0,1$, and $\langle \sigma_i ^2\rangle =1$ for $\sigma_i=\pm1$. The amplitude  $\alpha$ determines the strength of the pseudo-count. We expect $\alpha$ to vanish as $B\to\infty$, as regularization should not be necessary in the case of perfect sampling (assuming all the couplings in the underlying model for the data are finite). The example given above amounts to choosing $\alpha=\alpha^B=\frac 2{B+1}$.

\subsubsection{$L_1$- and $L_2$-norm regularization of couplings}
Another possibility to prevent couplings from being infinite is to consider a prior probability distribution for the couplings which discounts large coupling values. The log of the prior distribution then adds to the log-likelihood (\ref{llmf}), with the contribution
\begin{equation}\label{l2penality}
\Delta L(J) = -\frac {\gamma\;B} 2 \sum_{i<j} J_{ij}^2
\end{equation}
in the case of a Gaussian prior with variance $(\gamma\,B)^{-1}$ and mean zero. We have factored out $B$ in (\ref{l2}) to allow for a direct comparison with the log-likelihood of the data set (\ref{llmf}), which is proportional to $B$. In principle the prior should neither depend on the data, nor on their number, $B$. Hence $\gamma$ is expected to scale as $\frac 1B$, and to vanish in the case of perfect sampling. 
In the following we use the following $L_2$ regularization, instead of (\ref{l2penality}),
\begin{equation}\label{l2}
\Delta L(J) = -\frac {\gamma\;B} 2 \sum_{i<j} p_i\,(1-p_i) \,p_j \,(1-p_j)\, J_{ij}^2 \ .
\end{equation}
where $p_i=<\sigma_i>$  are the empirical averages of the binary variables. 
The reason is that the coupling matrix $\bf J$ maximizing the log-likelihood $L(c|J)+\Delta L(J)$ can be analytically calculated, see Eqs.~(21) and (35)  in \citep{Cocco:2012id} with ${\bf J} =Id -{\bf J}'$.

Another frequently used regularization scheme is $L_1$-norm regularization of the couplings, corresponding to a Laplacian prior distribution, which gives to the following additive contribution to the log-likelihood:
\begin{equation}\label{l1}
\Delta L(J) = -\gamma\;B \sum_{i<j} |J_{ij}|\ .
\end{equation}
$L_1$-norm regularization favors zero instead of small couplings, and produces sparse interaction graphs. In the following analysis we will see that the $L_1$-norm regularization is less adequate than the two other schemes presented here for mean-field inference.
No analytical expression exists for the optimal $J$. However, it can be found in a polynomial time using convex optimization techniques\citep{Friedman:2008tb}. 

\subsection{Case of the Potts model} \label{potts934}

The Potts model is a generalization of the Ising model in which each spin can take $q \ge 2$ values, hereafter called symbols, $a=1,2,\ldots, q$.  A mapping can be made onto the Ising model through the introduction of binary spins,  $\sigma_i^a=1$ if spin $i$ carries symbol $a$, and $\sigma_i^a=0$ otherwise.  Any $N$ Potts-spin configuration $\{a_i\}$ can be uniquely written as an $N\times q$ Ising-spin $\{\sigma_i^a\}$ configuration through this mapping. Reciprocally admissible Ising-spin configurations are such that the constraints
\begin{equation}\label{pottsconst}
\sum_a \sigma_i ^a =1 \ 
\end{equation}
hold for all sites $i$. In other words, at each site $i$, there is one and only one Ising spin equal to 1, with the remaining $q-1$ spins being equal to zero. 

The Hamiltonian of the Potts model may then be recast as an Ising Hamiltonian
\begin{equation}\label{pottsH}
H[\{a_i\}]\equiv  - \sum_i h_i (a_i) -\frac 12 \sum_{i \ne  j}
J_{i j} (a_i,a_j)= - \sum_i h_i (a)\, \sigma_i^a -\frac 12 \sum_{i \ne  j}
J_{ij} (a,b)\, \sigma_i^a \,  \sigma_j^b \ .
\end{equation}
Due to the constraints (\ref{pottsconst}) the local fields and the coupling parameters $h_i(a),J_{ij}(a,b)$ can be concomitantly changed, without affecting the Hamiltonian. One can check that $H$ is invariant under the change $J_{ij}(a,b)\to J_{ij}(a,b) + K_{ij} (b), h_i(b) \to h_i(b) - \sum_{j( \ne i)} K_{ij}(b)$ for arbitrary $K$. This invariance is called gauge invariance. In the following, we will restrict to one particular gauge, called zero-sum gauge, where for every pair of sites $(i,j)$, the sums of couplings along each column and each row of the $q\times q$ coupling matrix $J_{ij}(a,b)$ vanish.

The MF inference procedure presented in Section \ref{approxgauss} can be readily applied to the Ising representation of the Potts model. We obtain that the inferred coupling matrix, $J^{MF}_{ij}(a,b)$ is the pseudo-inverse of the correlation matrix, $c_{ij}(a,b)= \langle \sigma_i^a\sigma_j ^b\rangle - 
\langle \sigma_i^a\rangle \langle\sigma_j ^b\rangle $. The pseudo-inverse must be considered here in order to invert the correlation matrix in the $N(q-1)$-dimensional subspace orthogonal to the constraints (\ref{pottsconst}).

\section{Effects of regularization schemes: a toy-model analysis } \label{model}

In this Section we consider the effects of regularization through the analysis of very simple models with two or three spins only. For simplicity we will focus on the Ising model when exploring generic properties of regularized MF inference. Results particular to Potts models are discussed in Section~\ref{sec:potts} and \ref{sec:potts2}. 

In the following, we first infer the coupling between the spins using various regularization schemes and compare the outcome to the true value when sampling is perfect. We find that the MF approximation introduces errors into the inferred couplings, which can be corrected by strong regularization terms. In the Potts model case, the optimal regularization strength is found to depend on the number of symbols. We then investigate the effect of poor sampling on the pseudo-count performance. Finally, we consider the case of non-uniform couplings. Here we find that uniform $L_2$-norm regularization is suboptimal when couplings are strongly heterogeneous, and in this case the pseudo-count offers superior results compared to $L_2$-norm regularization.

\subsection{Ising models: Case of perfect sampling}

\subsubsection{Mean-field overestimates strong couplings}

\begin{figure}
\begin{center}
\includegraphics[width=6cm]{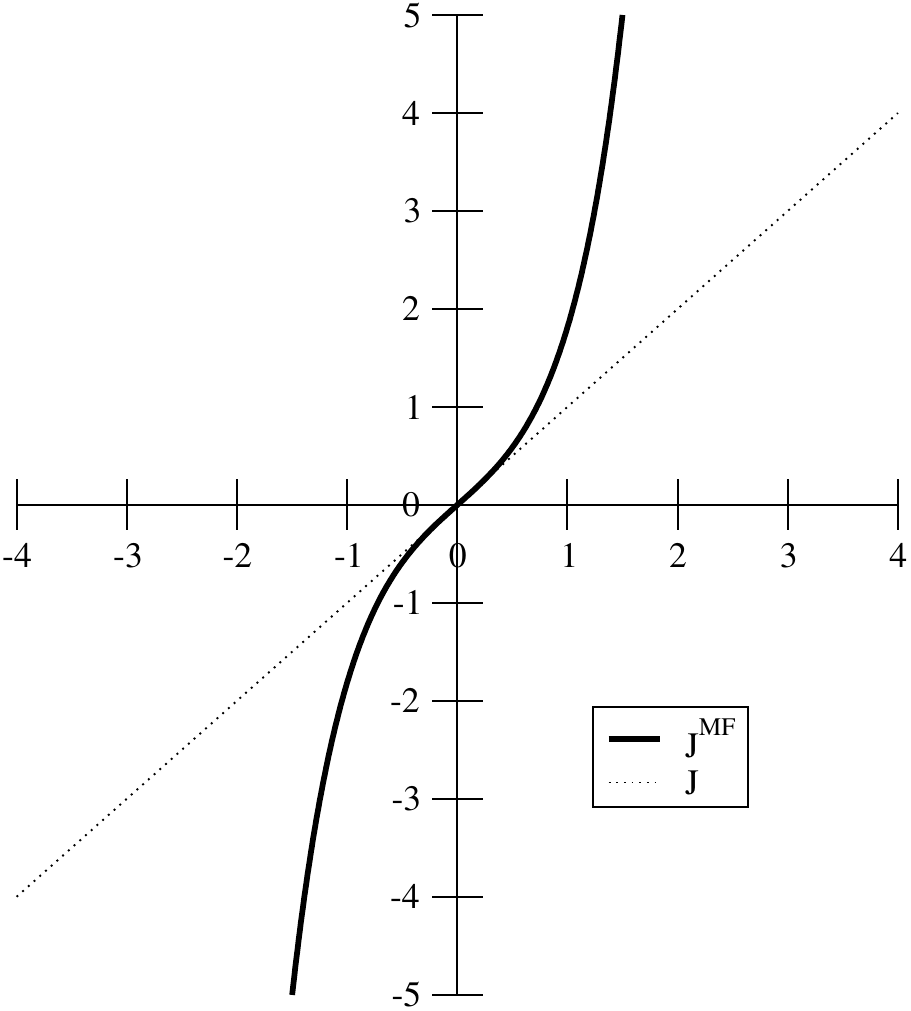}%2spins
\caption{Coupling $J^{MF}$ inferred with the mean-field approximation (\ref{mfj}) overestimates the true coupling value $J$ (dotted line) in a system of two spins when $J$ is large (in absolute value). }
\label{fig2spins}
\end{center}
\end{figure}

Let us consider two spins $\sigma_1,\sigma_2=\pm 1$ which are coupled through an interaction $J_{12}=J$, with no local fields ($h_1=h_2=0$). The magnetizations $\langle \sigma_1\rangle, \langle \sigma_2\rangle$ vanish, and the correlation is $c_{12}=\langle\sigma_1\sigma_2\rangle= \tanh J$. The $2\times 2$ correlation matrix therefore has elements unity on the diagonal, and $c_{12}$ on the off-diagonal. Using (\ref{inf}) we obtain the inferred coupling within MF approximation,
\begin{equation}\label{mfj}
J^{MF}(J)= \frac {c_{12}}{1-c_{12}^2}= \frac{\tanh J}{1- \tanh^2 J}\ .
\end{equation}
The MF prediction is plotted vs. the true coupling $J$ in Fig.~\ref{fig2spins}. The inferred coupling is in good agreement with the true value only for small couplings ($|J|<1$), and diverges very quickly, $|J^{MF}|\sim \frac 14 e^{2 |J|}$, as $J$ increases. The MF approximation drastically overestimates strong couplings.

\subsubsection{Strong regularizations with pseudo-count or $L_2$ correct for MF errors}

\begin{figure}
\begin{center}
{\bf \large A\hskip 6cm B \hskip 5.5cm C\hskip 6cm}\\
\includegraphics[width=5.5cm]{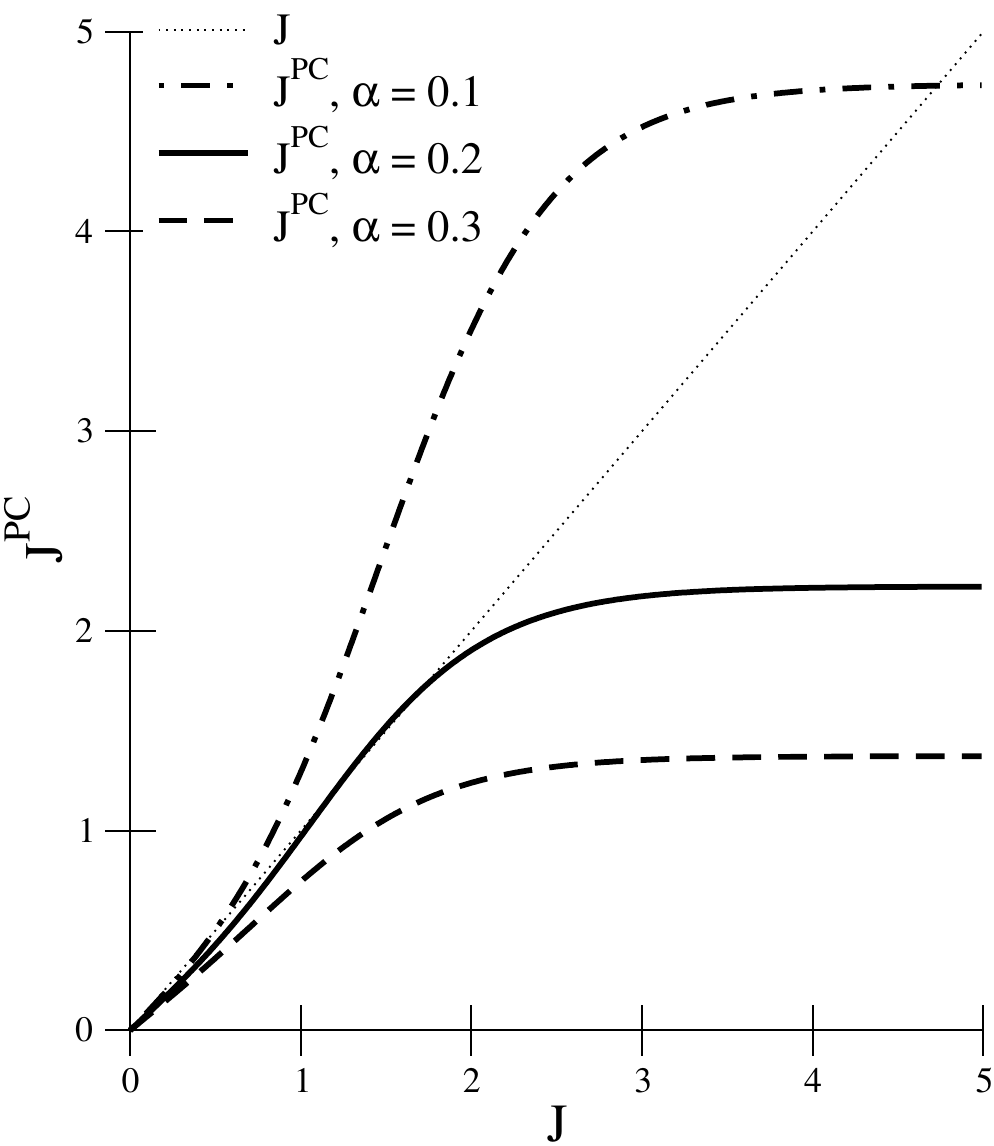}\hskip .4cm %2psinsPC
\includegraphics[width=5.5cm]{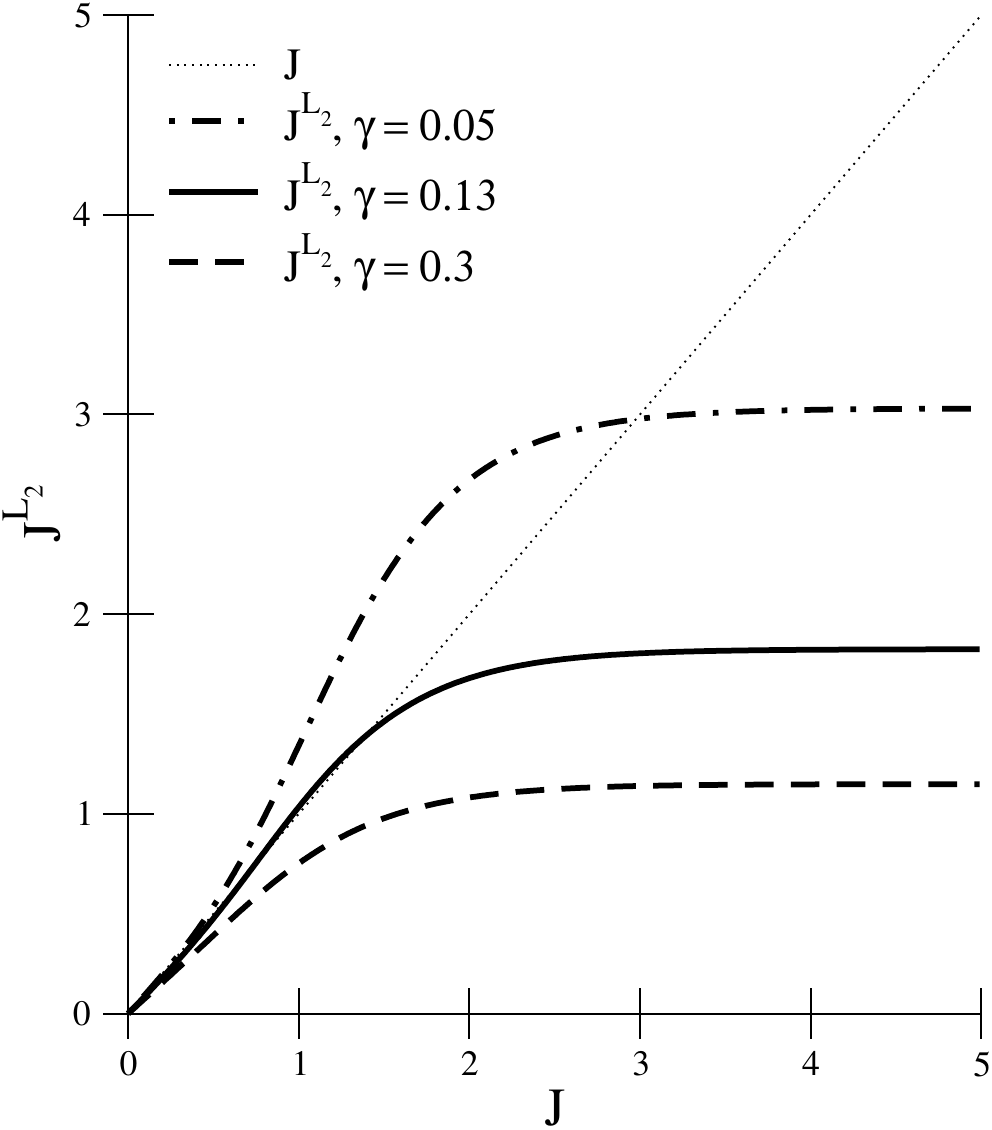}\hskip .4cm%2psinsL2
\includegraphics[width=5.5cm]{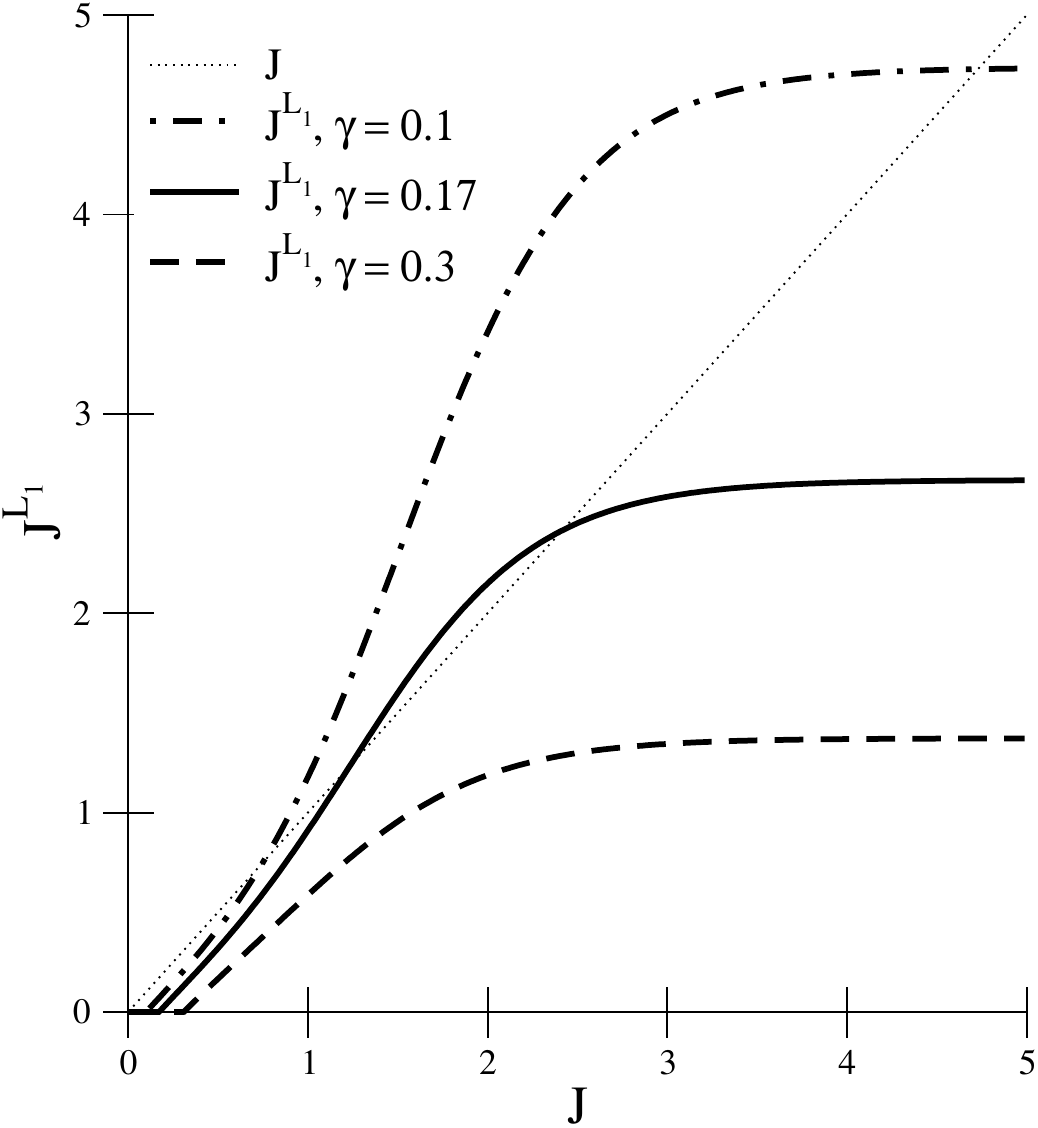}%2psinsL1
\caption{Strong regularization corrects for errors in the MF approximation for larger couplings, but $L_1$-norm regularization also introduces errors for small values of true couplings. Coupling inferred with the mean-field approximation with {\bf (A)} pseudo-count of intensity $\alpha$, $J^{PC}$, {\bf (B)}  with $L_2$-norm regularization of intensity $\gamma$, $J^{L_2}$, {\bf (C)} with the $L_1$-norm regularization of intensity $\gamma$, $J^{L_1}$, vs. true coupling value $J$ for a system of 2 spins. Due to the symmetry $J\to -J$ only the positive quadrant is shown.}
\label{fig2spinsregu}
\end{center}
\end{figure}

We start with the pseudo-count regularization of intensity $\alpha$, see (\ref{defpc2}). The inferred coupling is given by
\begin{equation}\label{pcj}
J^{PC}(J,\alpha)= \frac{(1-\alpha) \tanh J}{1- (1-\alpha)^2\tanh^2 J}\ .
\end{equation}
The MF prediction with pseudo-count (PC) regularization, $J^{PC}(J,\alpha)$, is plotted vs.~the true coupling $J$ in Fig.~\ref{fig2spinsregu}A for various values of $\alpha$. Unlike pure MF inference ($\alpha=0$), $J^{PC}$ saturates for very large couplings $J$ to a finite value, $J^{PC}(\infty,\alpha)=(1-\alpha)/(\alpha(2-\alpha))$. For intermediate couplings $|J|< J^{PC}(\infty,\alpha)$ we find that the agreement between $J^{PC}$ and $J$ is remarkably good for $\alpha\simeq 0.2$. Hence the pseudo-count with a large intensity (compared to the inverse of the number of data, which is infinite here since sampling is perfect) can correct for the dramatic overestimation of large couplings by the mean-field approximation. Couplings weaker in absolute value than the saturation value $J^{PC}(\infty,0.2)$ are precisely inferred, while larger couplings cannot be distinguished from $J^{PC}(\infty,\alpha)$.

A similar correction can be achieved with the $L_2$ regularization. Adding the penalty term (\ref{l2}) to the mean-field expression for the log-likelihood (\ref{llmf}) and maximizing over the $2\times 2$-coupling matrix $J$ we find that the off-diagonal coupling, $J^{L_2} (J,\gamma)$, is the root of the following implicit equation:
\begin{equation}
\tanh J = J^{L_2} (J,\gamma) \;\bigg( \gamma + \frac 2{1+\sqrt{1+4 \, J^{L_2} (J,\gamma)^2}}\bigg) \ .
\end{equation}
Fig.~\ref{fig2spinsregu}B compares the outcome to the true coupling for various values of $\gamma$. As in the pseudo-count case $J^{L_2}$ saturates for very large couplings $J$  to a finite and $\gamma$-dependent value, approximately equal to $1/\sqrt{2\gamma}$ for small $\gamma$. For intermediate couplings we find that the agreement between $J^{L_2}$ and $J$ is very good for $\gamma\simeq 0.13$. Hence $L_2$ regularization with a large intensity (again, compared to the inverse of the number of data, which is infinite here since sampling is perfect) avoids the divergence at large couplings introduced by the mean-field approximation, while being accurate for small coupling values. We observe that the coupling saturation value for the optimal $\gamma$ with the $L_2$ norm ($\simeq 1.8$) is however smaller than with the pseudo-count regularization $(\simeq 2.2$), compare Figs.~\ref{fig2spinsregu}A\&B; hence, the pseudo-count regularization offers an accurate inference over a slightly wider range of coupling values.

\subsubsection{$L_1$-norm regularization is less accurate than pseudo-count and $L_2$}

We now consider the $L_1$-norm regularization of intensity $\gamma$, obtained by adding the penalty term (\ref{l1}) to the log-likelihood (\ref{llmf}). An immediate calculation shows that the total log-likelihood is maximized by the coupling value
\begin{equation}
J^{L_1}(J,\gamma)= \left\{ \begin{array} {c c c}
\frac{\tanh J -\gamma}{1- (\tanh J -\gamma)^2} & \text{if} & \gamma \le \tanh J \ ,\\
0 & \text{if} & \gamma \ge  \tanh J \ . \end{array} \right. 
\end{equation}
The expression above for holds for positive $J$; for negative $J$ we have $J^{L_1}(J,\gamma)=-J^{L_1}(|J|,\gamma)$.
The inferred coupling $J^{L_1}(J,\gamma)$, is plotted vs.~the true coupling $J$ in Fig.~\ref{fig2spinsregu}C for various values of $\gamma$. As with pseudo-count and $L_2$  $J^{L_1}$ saturates for very large couplings $J$ to a finite value, $J^{L_1}(\infty,\alpha)=(1-\gamma)/(\gamma(2-\gamma))$. The novelty is that the inferred coupling vanishes for small $J$. Overall, for intermediate couplings $|J|< J^{L_1}(\infty,\alpha)$, the agreement between $J^{L_1}$ and $J$ is less precise than what can be achieved with the pseudo-count and $L_2$ regularization when, respectively, $\alpha$ and $\gamma$ are properly chosen. 
In \cite{Barton:2013fj}  the  inferred Ising couplings are compared with Gaussian couplings with  $L_1$ and $L_2$ norms  on neural data coming from multi-electrode recordings. Also on these data-sets,
in agreement with previous findings,  $L_2$  performs better over $L_1$ and  large regularization strengths are needed to improve the inference.

\subsection{Performance of pseudo-count as a function of the sampling quality}

We now focus on the pseudo-count scheme. We assume that a number, say, $B$, of configurations of the two spins are drawn at random, from the Ising model measure
\begin{equation}
P_J(\sigma_1,\sigma_2) = \frac {e^{J\sigma_1\sigma_2}}{2(e^J+e^{-J})} \ .
\end{equation}
The magnetizations $p_1\equiv\langle \sigma_1\rangle$ and $p_2\equiv \langle \sigma_2\rangle$, and the correlation $p_{12}\equiv \langle \sigma_{1}\sigma_2\rangle$ are then computed as empirical averages over the data. The joint probability density for these three quantities is given by 
\begin{eqnarray}
\rho \big(p_1,p_2,p_{12};B,J\big) &=& \sum_{0\le B_{++},B_{+-},B_{-+},B_{--}\le B}\binom{B}{B_{++},B_{+-},B_{-+},B_{--}} \frac {e^{J(B_{++}-B_{+-}-B_{-+}+B_{--})}}{(2(e^J+e^{-J}))^B}\nonumber \\
&\times & \delta \left(p_1- \frac {B_{++}+B_{+-}-B_{-+}-B_{--}}{B}\right)
\delta \left(p_2- \frac {B_{++}-B_{+-}+B_{-+}-B_{--}}{B}\right)\nonumber \\
&\times & \delta \left(p_{12}- \frac {B_{++}-B_{+-}-B_{-+}+B_{--}}{B}\right)\ .
\end{eqnarray}
We then define the average squared relative error on the inferred coupling as
\begin{equation}\label{er2}
\epsilon(B,J,\alpha) = \int_0^1 dp_1\,dp_2\, dp_{12}\; \rho \big(p_1,p_2,p_{12};B,J\big) \; \left( \frac {J^{PC}\big(p_1,p_2,p_{12},\alpha\big)}J-1\right)^2\ ,
\end{equation}
where
\begin{eqnarray}
J^{PC}\big(p_1,p_2,p_{12},\alpha\big)&=& J^{MF}\big( (1-\alpha) p_1, (1-\alpha) p_2, (1-\alpha) p_{12}+\alpha(1-\alpha) p_1p_2\big) \ , \nonumber \\
J^{MF}\big(p_1,p_2,p_{12})&=& \frac{p_{12}}{(1-p_1)(1-p_2)- p_{12}^2}\ ,
\end{eqnarray}
are the PC and MF predictions for the coupling given the magnetizations and the correlation. These expressions extend formulae (\ref{mfj}) and (\ref{pcj}) to the case of nonzero magnetizations.

Peculiar samples for which $J^{PC}$ and $J^{MF}$ diverge, {\em e.g.}~such that $p_1=p_2=0$ and $p_{12}=-1$,  may happen with nonzero (albeit exponentially small in $B$) probabilities. To get a well-defined and finite expression for the squared error in (\ref{er2}) we replace the term squared within parenthesis with the minimum of this term and one. The latter constant is arbitrary; any other choice would lead to $e^{-O(B)}$ changes to the error. With our choice, the relative square error cannot be larger than unity by construction.

\begin{figure}
\begin{center}
\includegraphics[width=9cm]{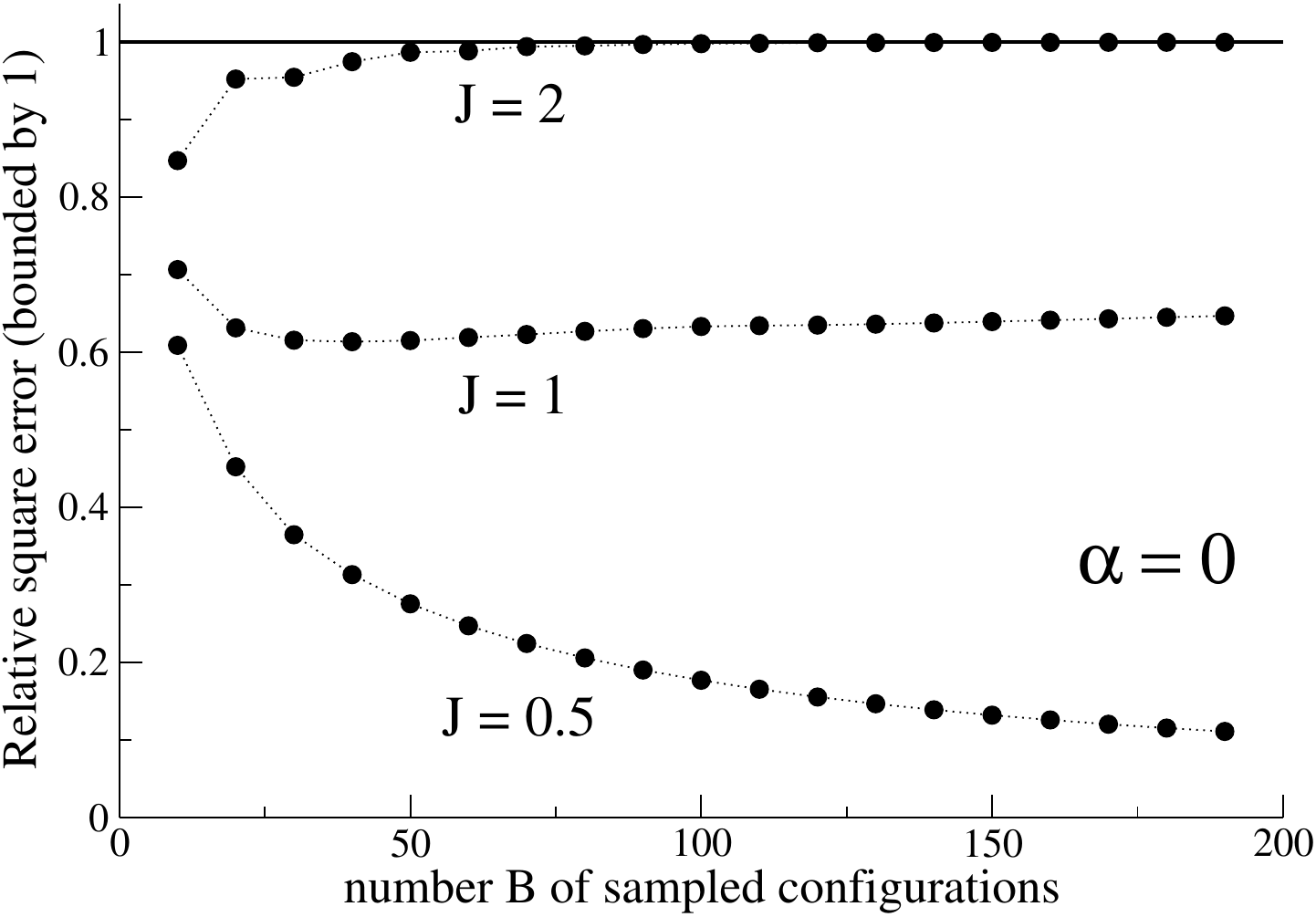}\vskip .5cm %2spinsalf0-eps-converted-to.pdf
\includegraphics[width=9cm]{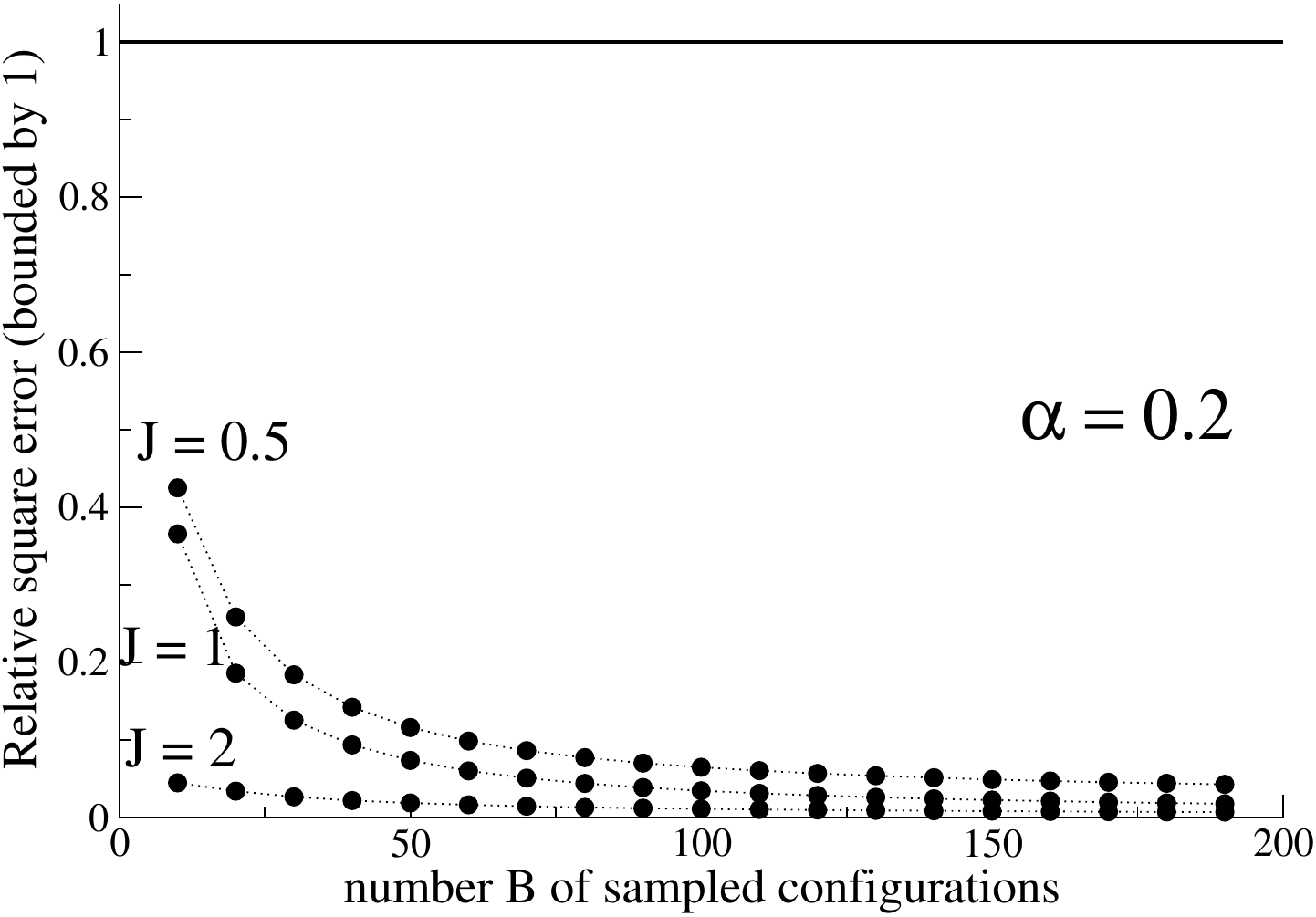}\vskip .5cm%2spinsalf02-eps-converted-to.pdf
\includegraphics[width=9cm]{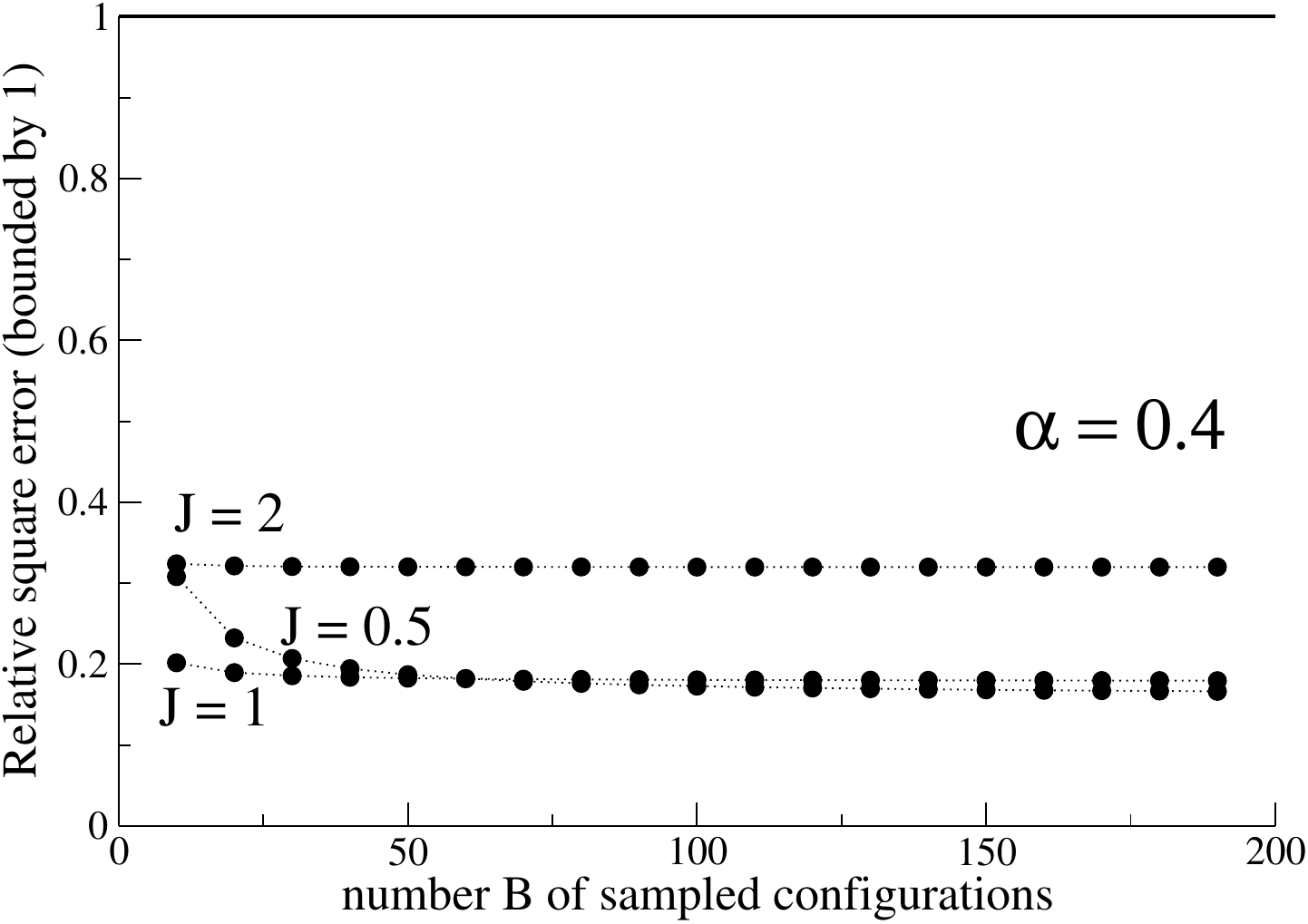}%2spinsalf04-eps-converted-to.pdf
\caption{Pseudo-count decreases the error on inferred couplings even when the couplings are weak, or sampling is large, but regularization that is too large leads to worse performance. Relative squared error $\epsilon(B,J,\alpha)$ (\ref{er2}) between the inferred coupling with pseudo-count of intensity $\alpha$ and the true coupling for a system of 2 spins as a function of the number $B$ of sampled configurations. Each panel corresponds to one value of $\alpha$, and each curve to one value of $J$.}
\label{fig2spinsalf0}
\end{center}
\end{figure}

We show in Fig.~\ref{fig2spinsalf0}(top) the relative square error as a function of $B$ in the absence of pseudo-count ($\alpha=0$). We observe that the error is small for small couplings $J$ if the number $B$ of configurations is large, and increases rapidly with $J$, in agreement with the findings of Fig.~\ref{fig2spinsregu}A. For the optimal value of the pseudo-count strength, $\alpha=0.2$, the relative error is a decreasing function of $B$, and saturates to a small value for all couplings, see Fig.~\ref{fig2spinsalf0}(middle); larger couplings produce larger correlations, and are easier to recover (require a smaller number of configurations) than smaller interactions. If the pseudo-count strength is too large, e.g.~$\alpha=0.4$, the error saturates to a finite and larger value, as shown in Fig.~\ref{fig2spinsalf0}(bottom). We conclude that the presence of a pseudo-count with fixed strength compensates the errors due to the mean-field approximation, even for a small number of sampled configurations.

\subsection{Efficiency of uniform regularization for non-uniform couplings} \label{sec-uniform}

An important question is whether a regularization scheme with uniform penalties, {\em i.e.}~equal to all pairs $(i,j)$ is appropriate in the case of a network of interactions with heterogeneous interactions. In order to study this point in a simple case we consider a system of three spins, with zero external fields and couplings $J_{12}=J_{13}\equiv \hat J_0$ and $J_{23}\equiv \hat J_1 \ne  \hat J_0$. We assume that sampling is perfect, and derive the spin-spin correlations $p_{ij}\equiv\langle \sigma_i\sigma_j\rangle$, with $1\le i<j\le 3$. We obtain
\begin{equation}
p_{12}=p_{13}= \frac{e^{2\hat J_1}\sinh(2\hat J_0)}{e^{2\hat J_1}\cosh(2\hat J_0)+1}\ , \quad p_{23}= \frac{e^{2\hat J_1}\cosh(2\hat J_0)-1}{e^{2\hat J_1}\cosh(2\hat J_0)+1}\ ,
\end{equation}
while all three magnetizations vanish.  

\subsubsection{Case of $L_2$-norm regularization}

We start with the $L_2$-norm regularization. We infer the three coupling values with MF by maximizing the log-likelihood $L^{MF}(c|J)$, where
\begin{equation}\label{j3by3}
J=\left(\begin{array}{ c c c} K_0 &J_0 &J_0\\J_0 &K_1 &J_1\\ J_0 &J_1 &K_1\end{array}\right)\ , \quad
c= \left(\begin{array}{ c c c} 1 &p_{12} &p_{12}\\p_{12} &1 &p_{13}\\ p_{12} &p_{13} &1\end{array}\right)
\ ,
\end{equation}
with an additive $L_2$-penalty term given by (\ref{l2}). In the following rather than fixing the values for $\hat J_0,\hat J_1$ and calculating the inferred couplings $J_0,J_1$ we do the opposite. The reason is that the maximization equations are complicated implicit equations over $J_0,J_1$ for given $\hat J_0,\hat J_1$, and are simpler to solve for $\hat J_0,\hat J_1$ given the values of $J_0,J_1$.

\begin{figure}
\begin{center}
{\bf \large A\hskip 8 cm B}\\
\includegraphics[width=8cm]{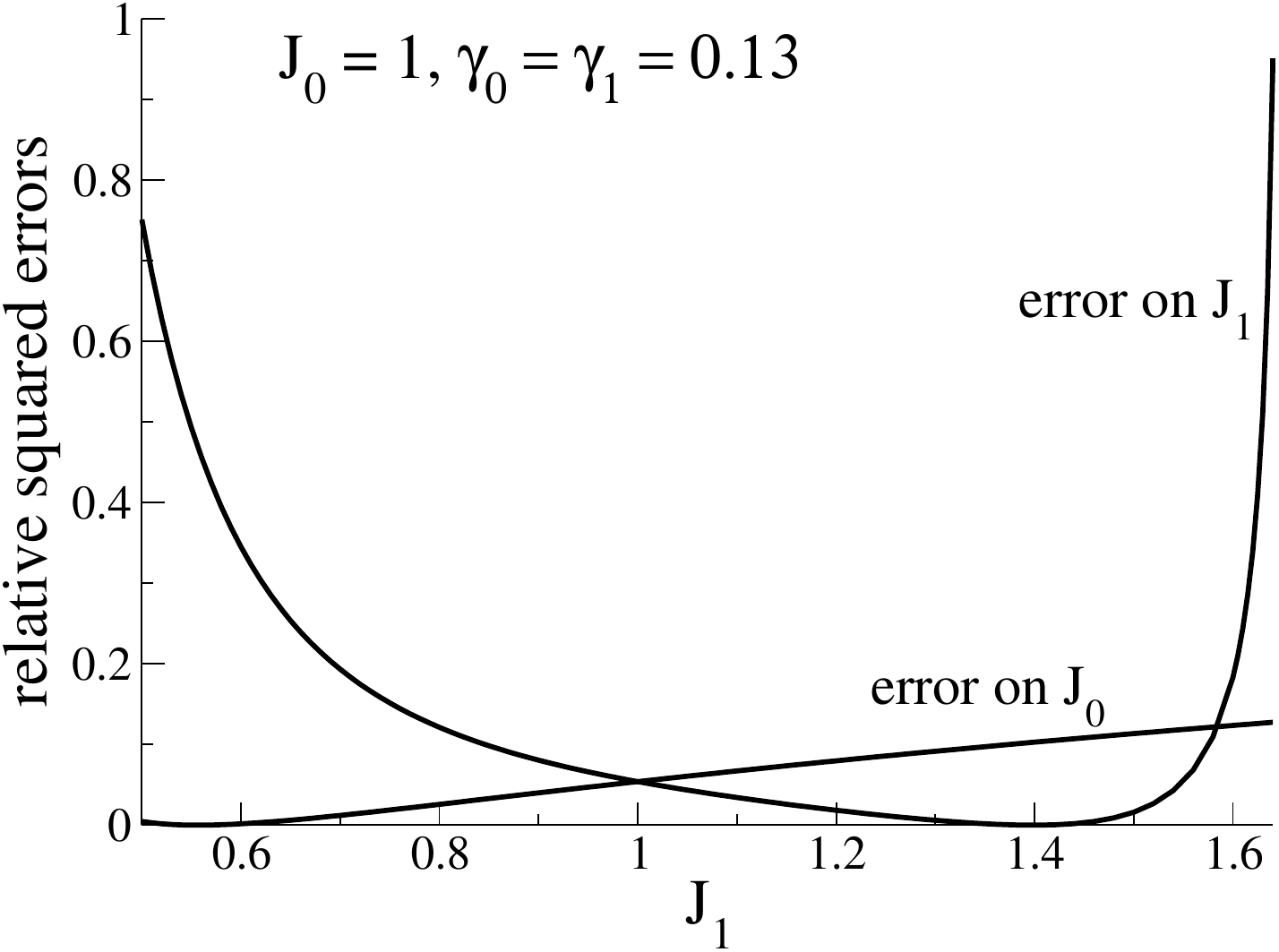}\hskip .5cm%error_jv_L2_gam013_j01-eps-converted-to.pdf
\includegraphics[width=8cm]{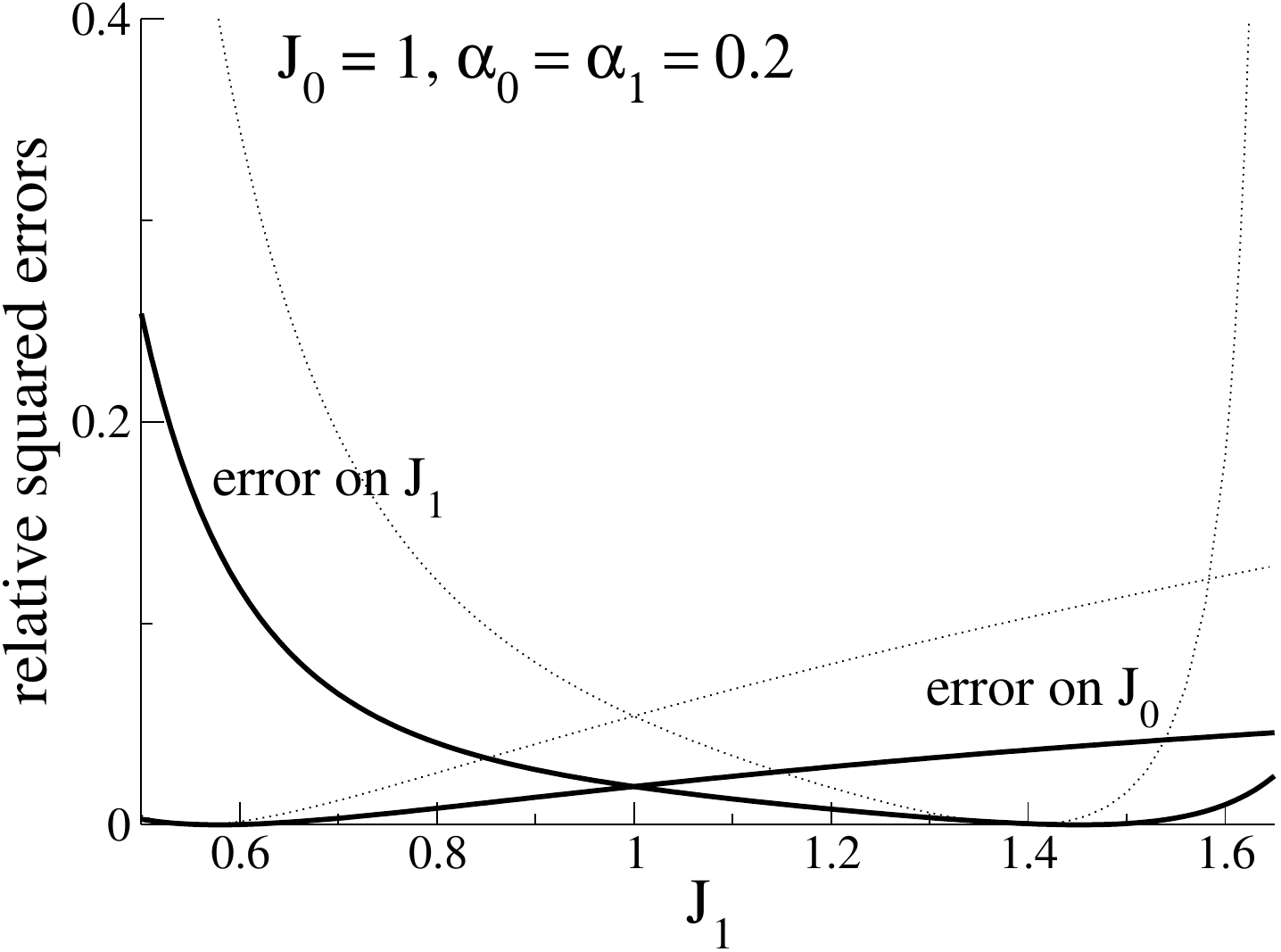}%error_jv_PC_alf02_j01-eps-converted-to.pdf
\caption{Relative squared error over the couplings with {\bf (A)} uniform $L_2$-norm penalty and {\bf (B)} uniform pseudo-count for a system of three spins, see text. The dotted lines in {\bf (B)} reproduce the relative squared errors found for the $L_2$-norm regularization in {\bf (A)}; Note the change in the vertical axis scale.}
\label{figtriangle}
\end{center}
\end{figure}

We show in Fig.~\ref{figtriangle}A the relative squared error between the true and inferred values for the couplings, as a function of $J_1$ for a fixed $J_0=1$. The value of the penalty is chosen to be $\gamma=0.13$, see Fig.~\ref{fig2spinsregu}B. We observe that the relative squared errors are small when $J_1$ and $J_0$ are close to each other as expected. However, when $J_1$ departs from the value of the other couplings, $J_0$, the relative error on $J_1$ becomes large. We conclude that imposing uniform $L_2$ penalties is not optimal for non-uniform couplings, and can lead to substantial errors in the inferred couplings.

\subsubsection{Case of pseudo-count}

We now consider the pseudo-count regularization. We infer the three coupling values with the MF approximation by maximizing the log-likelihood $L^{MF}(c|J)$, where $J$ is given by (\ref{j3by3}) and the regularized correlation matrix is
\begin{equation}
c= \left(\begin{array}{ c c c} 1 &(1-\alpha) p_{12} &(1-\alpha)p_{12}\\(1-\alpha)p_{12} &1 &(1-\alpha)p_{13}\\(1-\alpha) p_{12} &(1-\alpha)p_{13} &1\end{array}\right)
\ .
\end{equation}

We show in Fig.~\ref{figtriangle}B the relative squared error between the true and inferred values for the couplings, as a function of $J_1$ for a fixed $J_0=1$. The PC strength is chosen to be $\alpha=0.2$, see Fig.~\ref{fig2spinsregu}A. We observe that for a large range of values of $J_1$ compared to $J_0$ the errors on both couplings remains small. The accuracy is better than with the $L_2$-norm regularization.

\subsection{Potts model: Case of perfect sampling} 

\subsubsection{Homogeneous case} \label{sec:potts}

We now consider a two-spin Potts model, and denote by $q\ge 2$ the number of spin symbols. The model has Hamiltonian $H(a_1,a_2)=-J_0 \,\delta _{a_1,a_2}$, where $a_1$ and $a_2$ are the symbols corresponding to, respectively, spins 1 and 2. Each spin is equally likely to be in any of the $q$ symbols, and we will hereafter refer to this model as the homogeneous Potts model.

The Ising model is recovered when $q=2$; note that, as the difference between the energies of the equal-spin and different-spin configurations is $J$ in the Potts model, the corresponding Ising model coupling is $J/2$. In the following we will express the $q\times q$ coupling matrix $\bf J$ in a specific gauge, in which the sum of couplings over a row or a column of the matrix vanish
\begin{equation}\label{potts7unb}
{\bf J}= \left( \begin{array}{c c c c c}
J_A & J_B & J_B & \ldots & J_B \\
J_B & J_A & J_B & \ldots  & J_B \\
J_B & J_B & J_A & \ldots  & J_B \\
\ldots & \ldots & \ldots & \ldots & J_B \\
J_B & J_B & J_B & J_B & J_A
\end{array}\right) 
\quad \text{with} \quad 
J_A=\frac{q-1}q\, J_0 \ , \quad J_B=-\frac {J_0}q\ .
\end{equation}

The  two-spin correlations are
\begin{eqnarray}\label{potts1}
\langle \delta _{\sigma_1,a}\delta _{\sigma_1,a}\rangle &=&\langle \delta _{\sigma_2,a}\delta _{\sigma_2,a}\rangle =\frac 1q \ , \quad 
\langle \delta _{\sigma_1,a}\delta _{\sigma_1,b}\rangle =\langle \delta _{\sigma_2,a}\delta _{\sigma_2,b}\rangle = 0 \ , \nonumber \\
\langle \delta _{\sigma_1,a}\delta _{\sigma_2,a}\rangle &\equiv& p_{aa}=\frac{e^{J_0}}{q(e^{J_0}+q-1)} \ , \quad
\langle \delta _{\sigma_1,a}\delta _{\sigma_2,b}\rangle \equiv p_{ab}=\frac{1}{q(e^{J_0}+q-1)} \ .
\end{eqnarray}
where $a$ and $b$ denote different spin symbols. The $2q\times 2q$ entries of the connected correlation matrix $\bf C$ can be computed from those values, after subtraction of $1/q^2$.

\begin{figure}
\begin{center}
\includegraphics[width=12cm]{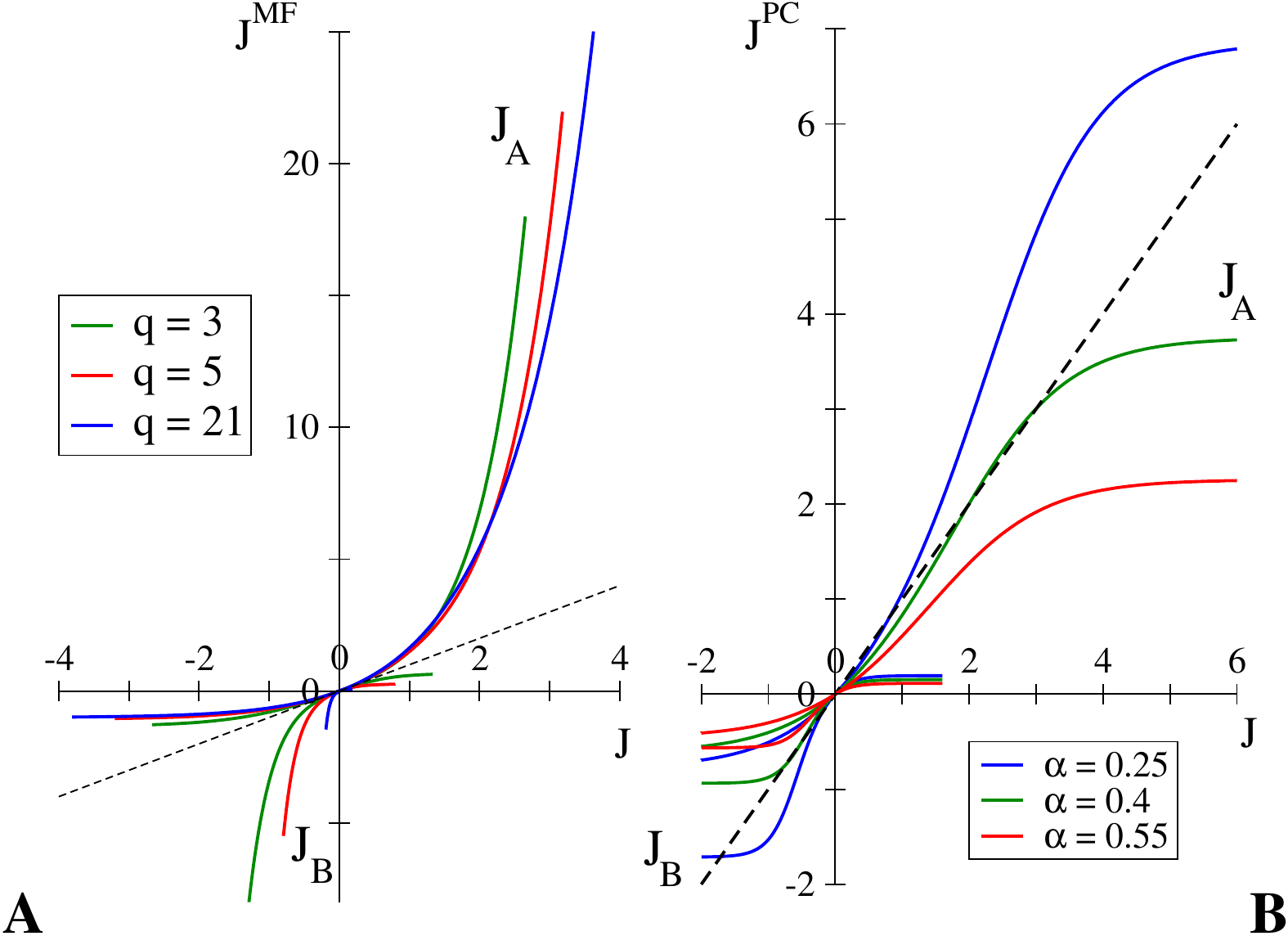}%2spins_potts
\caption{Homogeneous $q$-symbol Potts model. Couplings inferred with the mean-field approximation, $J_A^{MF}(J_0,q)$ (diagonal) and $J_B^{MF}(J_0,q)$ (off-diagonal) vs. true values coupling $J_A$ and $J_B$ in (\ref{potts7unb}). Curves were obtained through a parametric representation with $J_0$ running from -4 to 4; longer stretches would be obtained by increasing the range of values for $J_0$. The dashed line represents the $J^{MF}=J$ curve. ({\bf A}) No pseudo-count. ({\bf B}) With a pseudo-count of strength $\alpha$ (see values in the figure), and for $q=5$ symbols.}
\label{fig2spinspotts}
\end{center}
\end{figure}

Note that the sum of the elements of $\bf C$ over a line or a column is equal to zero. This property of $\bf C$ reflects the fact that each spin takes one symbol value. To obtain the mean-field prediction for the coupling, ${\bf J}^{MF}$, we consider minus the pseudo-inverse of $\bf C$. Again, the sum of the elements of ${\bf J}^{MF}$ over a line or a column is equal to zero. The inferred Potts couplings $J_A^{MF}$ and $J_B^{MF}$ correspond to, respectively, the diagonal and off-diagonal entries of the off-diagonal $q\times q$ blocks of ${\bf J}^{MF}$. Some simple algebra gives:
\begin{equation}\label{mfpotts}
J_A^{MF}(J_0,q)= \frac{ q(q-1) (p_{aa}-p_{ab})}{1- q^2 (p_{aa}-p_{ab})^2} \ ,
\quad J^{MF}_B(J_0,q)= -\frac{ q (p_{aa}-p_{ab})}{1- q^2 (p_{aa}-p_{ab})^2} \ ,
\end{equation}
where the correlations $p_{aa}$ and $p_{ab}$ are given in (\ref{potts1}). The MF inferred couplings are shown as functions of $J_0$ for different $q$ in Fig.~\ref{fig2spinspotts}A.  For each value of $q$ there are two branches corresponding to $J_A$ and $J_B$. The upper branch $J_A$ coincides with the lower branch $J_B$ after rescaling of $J$ and $J^{MF}$ by the factor $-1/(q-1)$. As in the Ising case, the MF prediction is quantitatively correct for weak couplings, but strongly overestimates the right coupling value for large $J$ (in absolute value).  Contrary to the Ising case there is an asymmetry between the positive and negative values of $J$ (for $q\ge 2$) along each branch $J_A$ or $J_B$. 

In the presence of a pseudo-count of strength $\alpha$ the correlation functions $p_{aa}$ and $p_{ab}$ are multiplied by $(1-\alpha)$. We can use again formula (\ref{mfpotts}) to obtain the corresponding couplings, which we denote by $J_A^{PC}$ and $J_B^{PC}$. Results for $q=5$ symbols and three values of the pseudo-count, ranging from $\alpha=0.25$ to 0.55, are shown in Fig.~\ref{fig2spinspotts}B. As in the Ising case we find that the inferred coupling $J^{PC}$ saturates to a finite value when $J\to \pm\infty$. There is an optimal value of the pseudo-count strength $\alpha$ such that the inferred and true coupling values are close to one another for positive $J_A$, and negative $J_B$. We observe, however, that for negative $J_A$ and positive $J_B$, the pseudo-count is not able to correct the errors produced by the MF approximation.

\begin{figure}
\begin{center}
\includegraphics[width=8cm]{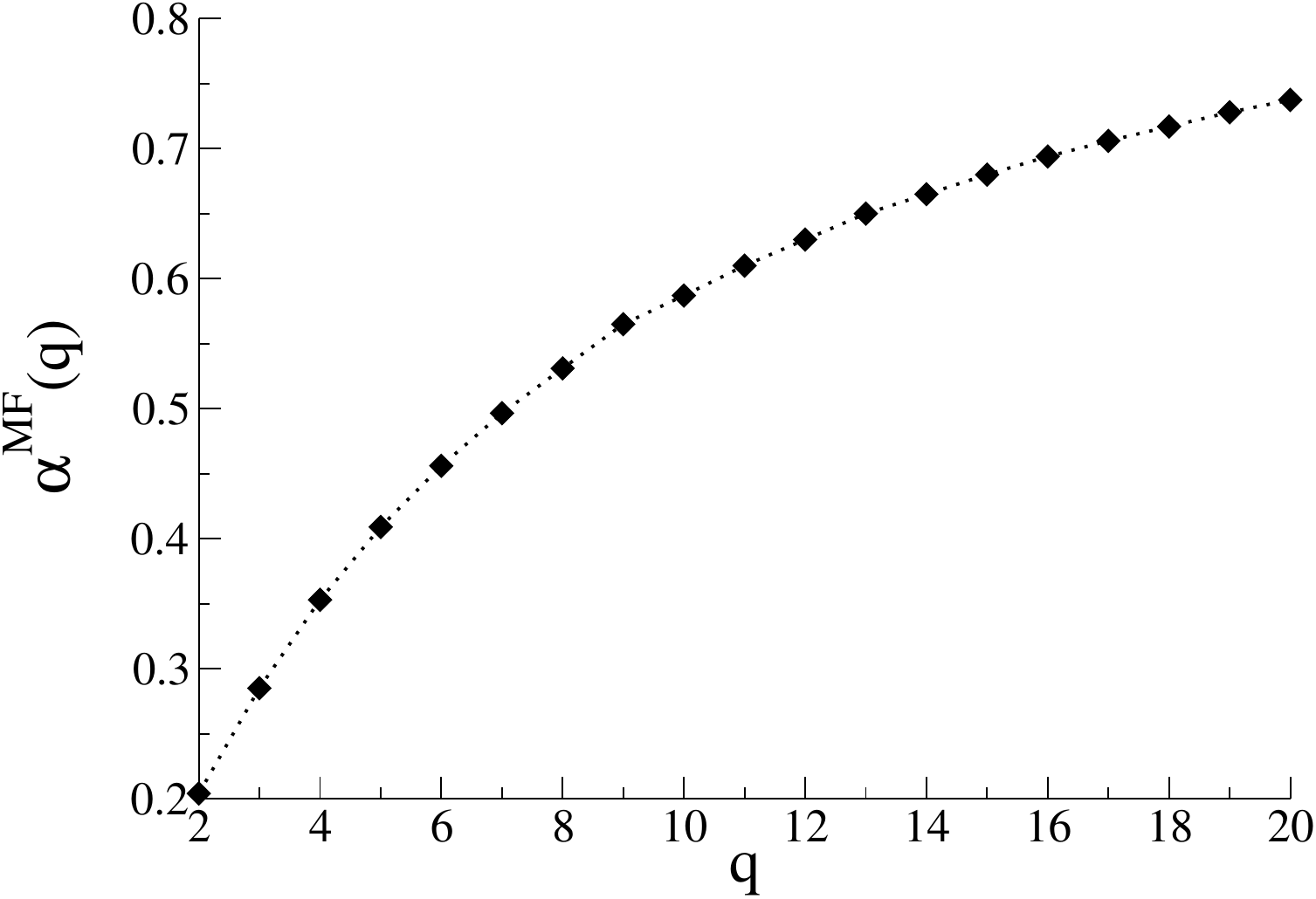}%alphaq
\caption{Optimal pseudo-count strength $\alpha^{MF}(q)$ as a function of the number of Potts symbols, $q$. The dotted line serves as a guide to the eye. }
\label{alphaq}
\end{center}
\end{figure}

We may define the optimal pseudo-count $\alpha^{MF}(q)$ as the largest value of $\alpha$ such that $J^{PC}=J$ has a nonzero solution (for positive $J_A$ or, equivalently, for negative $J_B$). In other words, when $\alpha=\alpha^{MF}(q)$, the representative curve for the inferred coupling touches the $J^{PC}=J$ line tangentially (dotted line in Fig.~\ref{fig2spinspotts}B). The value of $\alpha^{MF}(q)$ is shown as a function of $q$ in Fig.~\ref{alphaq} for $q$ ranging between 2 and 20. We observe a monotonic increase of the optimal pseudo-count with $q$, from $\alpha\simeq 0.2$ for $q=2$ to $\simeq 0.74$ for $q=20$. Our finding is in good agreement with empirical works on protein covariation, where the pseudo-count is often taken to be 0.5, but may vary between 0.3 and 0.7 depending on the protein family under consideration. 

\subsubsection{ Heterogeneous case} \label{sec:potts2}

We will now study a simple heterogeneous case, in which one of the $q$ symbols, say $a=1$, has a larger frequency than the other symbols. To do so we consider the Hamiltonian $H(a_1,a_2)=-J_0\,\delta _{a_1,a_2} - J_1\, \delta_{a_1,1}(1-\delta_{a_2,1})$ for the two spin symbols on the two sites. The coupling parameter $J_1$ acts as a field along the $a=1$ direction. The frequency $p_i^1$ of the first symbol, $a=1$, is the same on both sites and is larger than $\frac 1q$ if the $J_1>0$, and smaller than $\frac 1q$ if $J_1<0$. All other symbols $a=2,\ldots, q$ are equally likely with a frequency $ \frac{1-p_i^1}{q-1}$.  

In the zero-sum gauge, in which the sums of couplings along each row and each column vanish, the coupling matrix reads
\begin{equation}\label{potts7}
{\bf J}= \left( \begin{array}{c c c c c c}
J_A & J_B & J_B & J_B & \ldots & J_B \\
J_B & J_C & J_D & J_D & \ldots  & J_D \\
J_B & J_D& J_C & J_D & \ldots  & J_D \\
J_B & J_D& J_D & J_C & \ldots  & J_D \\
\ldots & \ldots & \ldots & \ldots & \ldots & J_D \\
J_B & J_D & J_D & J_D &J_D  & J_C
\end{array}\right) 
\quad \text{with} \quad \left\{ \begin{array}{c c c}
J_A&=&\frac{q-1}q\, J_0 - \frac{(q-1)^2}{q^2}\,J_1 \ , \\
J_B&=&-\frac{J_0}q\, + \frac{(q-1)}{q^2}\, J_1\ ,\\
J_C&=&\frac{q-1}q\, J_0 - \frac{J_1}{q^2} \ ,\\
J_D&=&-\frac{J_0}q- \frac{J_1}{q^2} \ .
\end{array}\right.
\end{equation}

The off-diagonal $q\times q$ block of minus the pseudo-inverse of the correlation matrix is the coupling matrix ${\bf J}^{MF}$ within the MF approximation, which fulfills the same gauge condition as ${\bf J}$. We obtain four couplings $J^{MF}_L(J_0,J_1,q)$, which can be compared to the four couplings $J_L(J_0,J_1,q)$ defined in (\ref{potts7}), with $L=A,B,C,D$. The homogeneous case, studied in the previous section, is recovered when $J_1=0$. In this case we have degenerate couplings: $J_A=J_C$ and $J_B=J_D$. 

\begin{figure}
\begin{center}
\includegraphics[width=14cm]{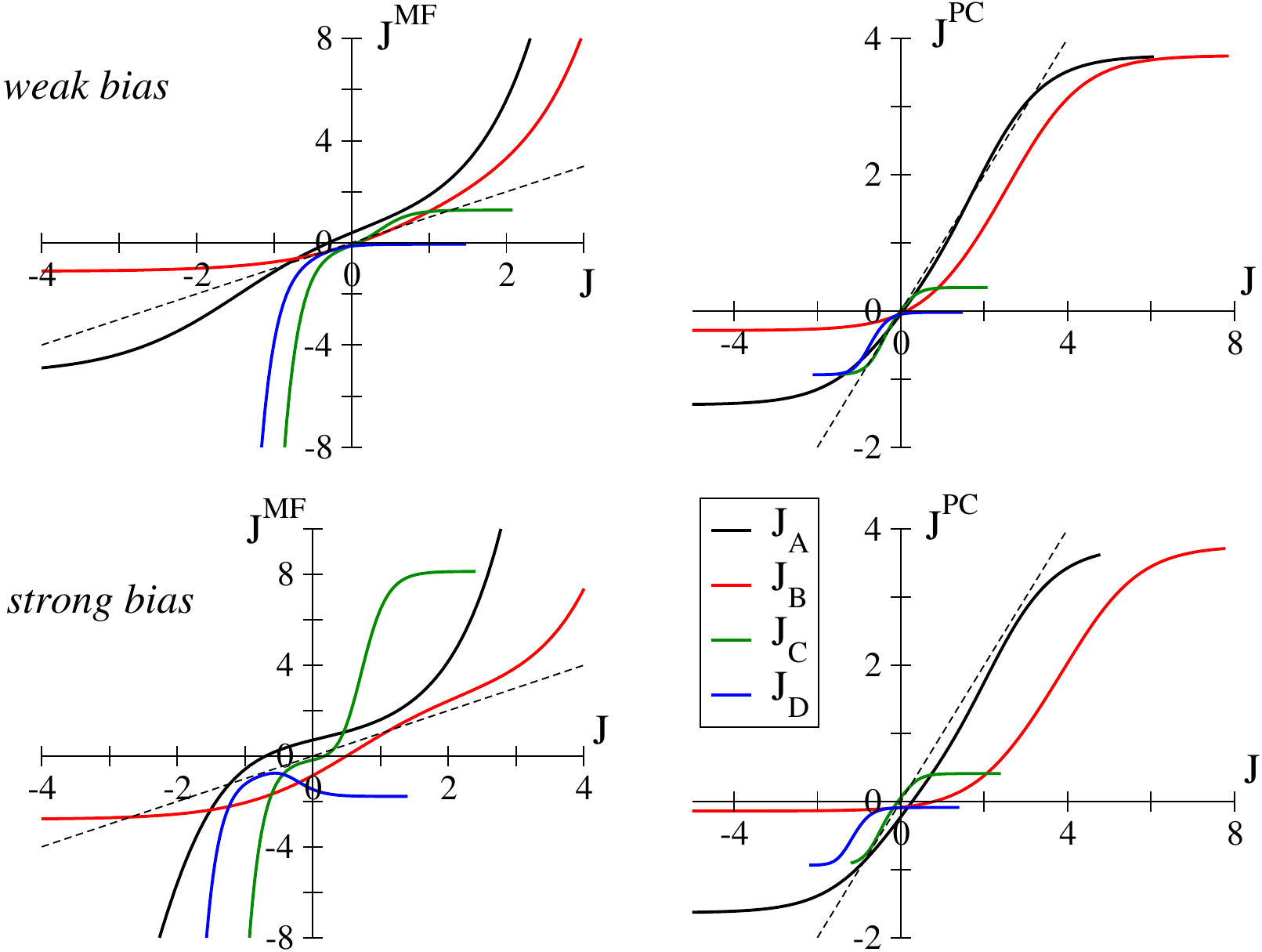}%potts9
\caption{Heterogeneous $q$-symbol Potts model. Couplings inferred with the mean-field approximation, $J^{MF}_L(J_0,J_1, q=5)$, (left) and with a pseudo-count $J_L^{PC}(J_0,J_1,q=5,\alpha=0.4)$ (right), vs. true couplings $J_L(J_0,J_1,q)$; $L=A,B,C,D$ labels the four different branches. The value of the pseudo-count strength $\alpha=0.4$, has been chosen according to Fig.~\ref{alphaq}. Curves were obtained through a parametric representation with $J_0$ running from -8 to 8; longer stretches would be obtained by increasing the range of values for $J_0$. Top: 'weak' bias ($J_1=3$), bottom: 'strong' bias  ($J_1=5$).}
\label{figpotts9}
\end{center}
\end{figure}
 Figure~\ref{figpotts9} shows the couplings inferred with the MF approximation against their true values for the $q=5$-symbol Potts model, in a 'weakly' biased case (top row, corresponding to $J_1=3$) and in a 'strongly' biased case (bottom row, corresponding to $J_1=5$). Note that the terms 'weak' and 'strong' have no absolute meaning here, as the bias is not constant when $J_0$ varies. In the weak bias case, we observe that the degeneracy between the couplings is lifted, compare with the $q=5$ curves in Fig.~\ref{fig2spinspotts}A. As the bias gets stronger (Fig.~\ref{figpotts9}, bottom \& left) the branches corresponding to $J_B$ and $J_D$ show markedly different behavior; for instance, the discrepancy between the inferred and true values of the couplings $J_B$ may largely exceed the errors on the diagonal couplings $J_A$ and $J_C$. Varying the value of $q$ does not qualitatively affect the results above, with the exception that the off-diagonal couplings $J_B,J_D$ become smaller as $q$ grows, in agreement with Fig.~\ref{fig2spinspotts}A.

The introduction of a large pseudo-count corrects, to some extent, the errors resulting from the MF approximation, see Fig.~\ref{figpotts9}, right panels. Two remarks can be made. First, the quality of the inference is better for the couplings corresponding to the symbols with high probability, here $J_A$ and $J_B$, as $\sigma_1=1$ is the most frequent symbol for $J_1>0$, see (\ref{potts7}). Secondly, when the bias increases, the quality of the inference does not decrease much for the couplings associated to frequent symbols ($J_A,J_B$), but strongly deteriorates for the other couplings ($J_C,J_D$). As a consequence, the inferred couplings occupy a larger part of the $J^{PC}<J, J>0$ and $J^{PC}>J, J<0$ portions of the $(J,J^{PC})$ plane.

\section{Numerical simulations for large systems}

To better understand mean-field inference on larger, more realistic data sets, we have tested the accuracy of the interaction graph recovered by pseudo-count and $L_2$-regularized MF inference for a variety of Ising (Section \ref{isingsimu}) and Potts models (Section \ref{pottssimu}). 

\subsection{Results for the Ising model}\label{isingsimu}

\begin{figure*}
\begin{center}
\includegraphics*[width=7.0in]{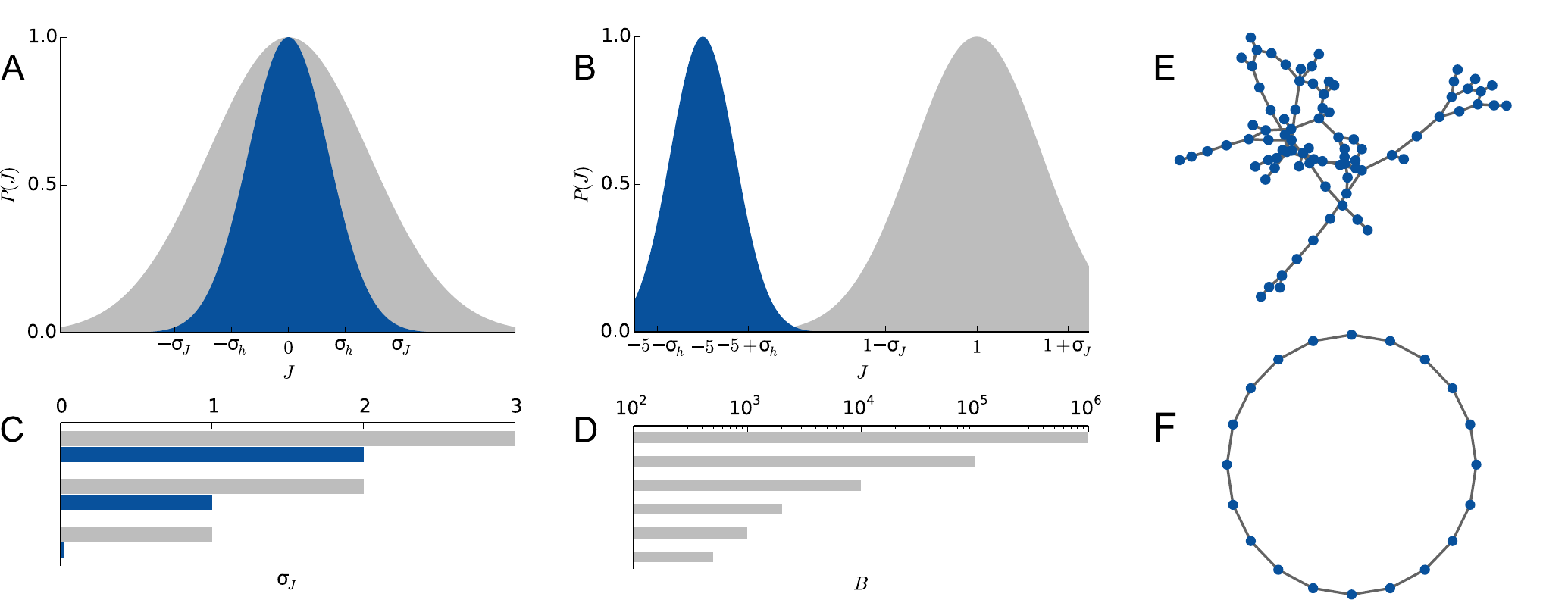}%{parameter_plot}
\caption{
Representation of tested model parameters.  
Fields and nonzero couplings were selected according to model (\textbf{A}), all fields and couplings normally distributed with means $\bar{h}=0$, $\bar{J}=0$ and standard deviations $\sigma_h$, $\sigma_J$, respectively, or model (\textbf{B}), strong negative fields and couplings normally distributed with means $\bar{h}=-5$, $\bar{J}=1$ and standard deviations $\sigma_h$, $\sigma_J$. Distributions for fields are shaded \textit{dark}, values for couplings are \textit{light}.
(\textbf{C}) For each model a range of $\sigma_h$ (\textit{dark}) and $\sigma_J$ (\textit{light}) was tested.
(\textbf{D}) Correlations used for the MF inference were computed using $B$ samples from a Monte Carlo simulation of the model, with $B$ tested over a range from $500$ to $10^6$.
All permutations of the above parameters were considered for each choice of the network topology: (\textbf{E}) Erd\H{o}s-R\'enyi graph where edges are kept with probability $p=2/N$ or $p=4/N$, and (\textbf{F}) $1D$ lattice with nearest neighbor couplings.
In all cases we take the system size $N=100$.
\label{parameters}}
\end{center}
\end{figure*}

We have tested Ising models with different network topologies, random distributions of the couplings, and using differing numbers of samples to compute the correlations (Fig.~\ref{parameters}). All the simulations reported in this Section were performed with spins taking values $0,1$.

The accuracy of the MF inference was quantified in two ways. First, we considered the standard root mean square (RMS) error between the inferred couplings $J^{inf}$ and those in the true model $J^{true}$,
\begin{equation} \label{RMSJ}
\Delta_J = \sqrt{\frac{2}{N (N-1)}\sum_{i<j} \left(J^{inf}_{ij}-J^{true}_{ij}\right)^2} \, , 
\end{equation}
where $N$ is the system size. The RMS error captures the absolute difference between the true and inferred couplings, but is unable to clearly distinguish whether the \emph{relative} ordering of the couplings has been correctly inferred. This limitation is problematic since many practical applications, such as the prediction of protein contacts from MF inference on sequence data \citep{Morcos:2011ct, Marks:2011hj, Hopf:2012kv}, rely on proper rank ordering of the inferred couplings rather than their absolute magnitude.

Information about the correct rank ordering can be determined from the rank correlation between the true and inferred couplings. To do this, we assigned each true coupling a rank according to its absolute value, with the largest coupling assigned rank 1, and the smallest rank $n_{\rm NZ}$, where $n_{\rm NZ}$ is the total number of nonzero couplings. All couplings exactly equal to zero are simply assigned rank $n_{\rm NZ}+1$. We then computed the Pearson correlation $\rho_J$ between the rank of the top $n_{\rm NZ}$ inferred couplings and their true counterparts, measuring how well the ordering of the top inferred couplings matches the true ordering in the underlying model. Letting $\{i_k,j_k\}$, with $k=1,\ldots,n_{\rm NZ}$, denote the pair indices of the largest $n_{\rm NZ}$ inferred couplings, this is
\begin{equation} \label{rhoJ}
\rho_J = \frac{1}{\sigma_{r(J^{true})} \, \sigma_{r(J^{inf})}} \sum_{k=1}^{n_{\rm NZ}} \left(k - \frac{n_{\rm NZ}+1}{2}\right) \, \Big(r(J_{i_k j_k}^{true}) - \bar{r}^{true}\Big) \, .
\end{equation}
Here $r(J)$ is the rank of coupling $J$, $\bar{r}$ the average rank, and $\sigma_{r(J)}$ the standard deviation of the ranks. Note that, since the inferred couplings are ranked from $1$ to $n_{\rm NZ}$, their average rank $\bar{r}^{inf} = (n_{\rm NZ} + 1)/2$. The fraction $R$ of true nonzero couplings included within the top $n_{\rm NZ}$ inferred couplings was also computed. In this way, we can assess how well the inferred couplings recover real couplings from the underlying model, and the degree to which their relative ordering is preserved.

\subsubsection{Regularization improves the quality of mean-field inference: an example on a $1D$ lattice}

\begin{figure*}
\begin{center}
\includegraphics*[width=.8\columnwidth]{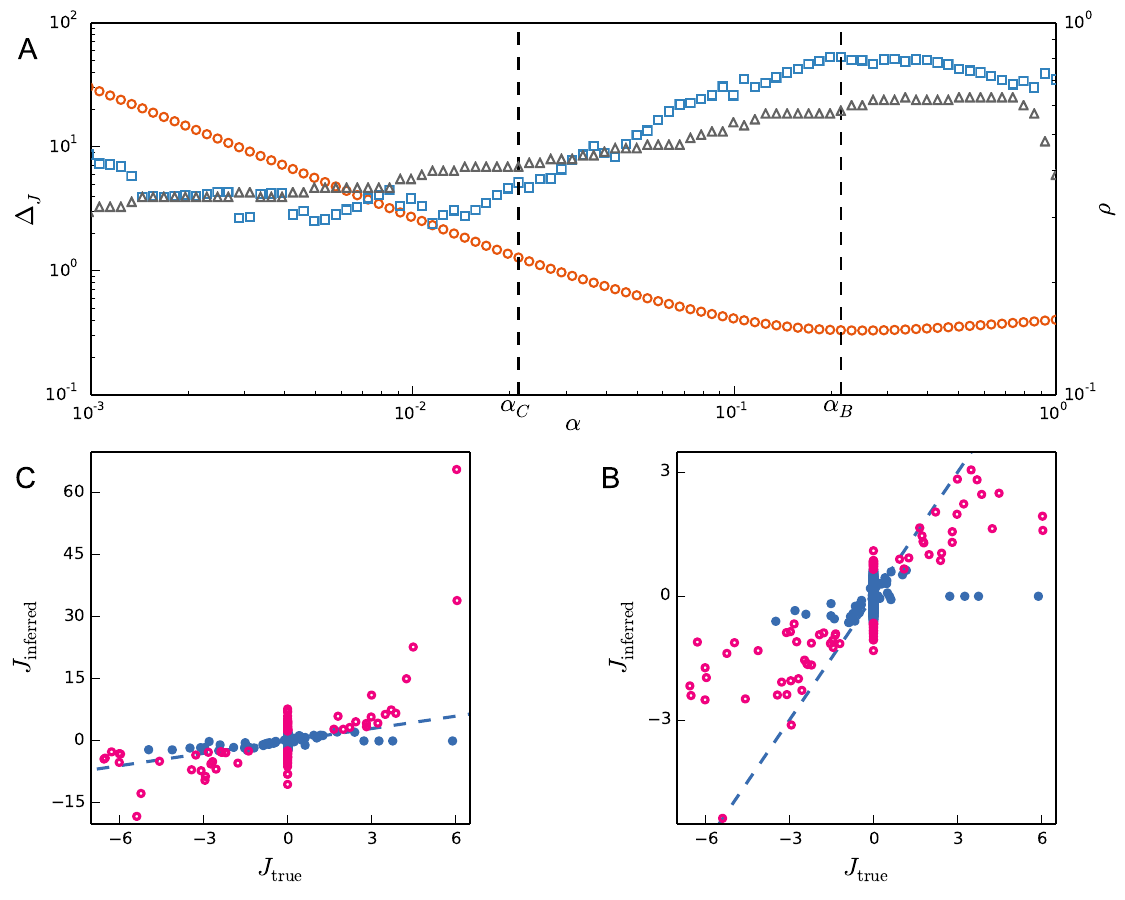}%1d-hmu0-jmu0-jsig3-B500-pc-JMerged
\caption{
Typical example trajectory of inference quality as a function of the regularization strength for a single randomly chosen set of couplings and fields. At larger regularization strengths the inferred couplings are similar in magnitude to the true couplings, but at lower regularization strengths the value of the strongest inferred couplings begins to diverge.
(\textbf{A}) RMS error $\Delta_J$ \ref{RMSJ} (\textit{circles, left axis}), rank correlation $\rho_J$ \ref{rhoJ} (\textit{squares, right axis}), and fraction of nonzero couplings recovered $R$ (\textit{triangles, right axis}), as a function of the pseudo-count $\alpha$.
(\textbf{B}) Comparison of true and inferred couplings at $\alpha=\alpha_B\approx0.2$, where performance is maximized. This value of the pseudo-count agrees well with the optimal pseudo-count for the Ising model $\alpha^{MF}=0.2$ shown in Fig.~\ref{alphaq}. Largest couplings are denoted by open circles, others are denoted by closed circles. A dashed line marks the $J^{inf}=J^{true}$ line.
(\textbf{C}) Comparison of true and inferred couplings at $\alpha=\alpha_C\approx0.02$.
The true model is a $1D$ spin chain, with zero fields and couplings normally distributed with mean zero and standard deviation $\sigma_J=3$. $B=500$ Monte Carlo samples were used to compute the correlations used for the MF inference.
\label{ex-trajectory}}
\end{center}
\end{figure*}

As a typical example, in Fig.~\ref{ex-trajectory} we show the performance of the mean-field inference as a function of pseudo-count for a model system with nearest-neighbor interactions on a $1D$ lattice and with a poor sampling on only $B=500$ configurations. As the pseudo-count is lowered from its maximum at $\alpha=1$, RMS error $\Delta_J$ (\ref{RMSJ}) decreases and the rank correlation $\rho_J$ (\ref{rhoJ}) improves until a peak is reached at $\alpha \simeq 0.2$ (Fig.~\ref{ex-trajectory}A), in excellent agreement with the optimal value of the pseudo-count strength necessary to correct the MF approximation, $\alpha^{MF}=0.2$  theoretically found for the  Ising system with two spins (Fig.~\ref{alphaq}). 
%For , $\alpha <\alpha^{MF}$  the RMS error rises sharply, and rank correlation decreases. 
At $\alpha\simeq\alpha^{MF}$ the largest true couplings are recovered well, and the inferred couplings are similar to the true ones in magnitude (Fig.~\ref{ex-trajectory}B). At lower values of the pseudo-count the largest inferred couplings are much larger than their true counterparts, and couplings that are zero in the true model are more likely to be inferred as large, causing performance to degrade (Fig.~\ref{ex-trajectory}C).
As the pseudo-count strength is decreased further   to   $\alpha\simeq\alpha^B=1/B$, the rank correlation $\rho_J$ and fraction of nonzero couplings recovered $R$ reach a plateau.
% This scale provides a natural 
%a natural cutoff, below which the effects of the pseudo-count regularization become small with respect to the size of the correlations themselves, which are $O(1/B)$ at smallest for finitely sampled data.
Below this scale the RMS error $\Delta_J$ can continue to rise, as couplings singular in the limit $\alpha\to 0$ become progressively larger.

\subsubsection{Scaling of the optimal regularization strength with sampling depth and coupling strength on random graphs}

\begin{figure*}
\begin{center}
\includegraphics*[width=3.5in]{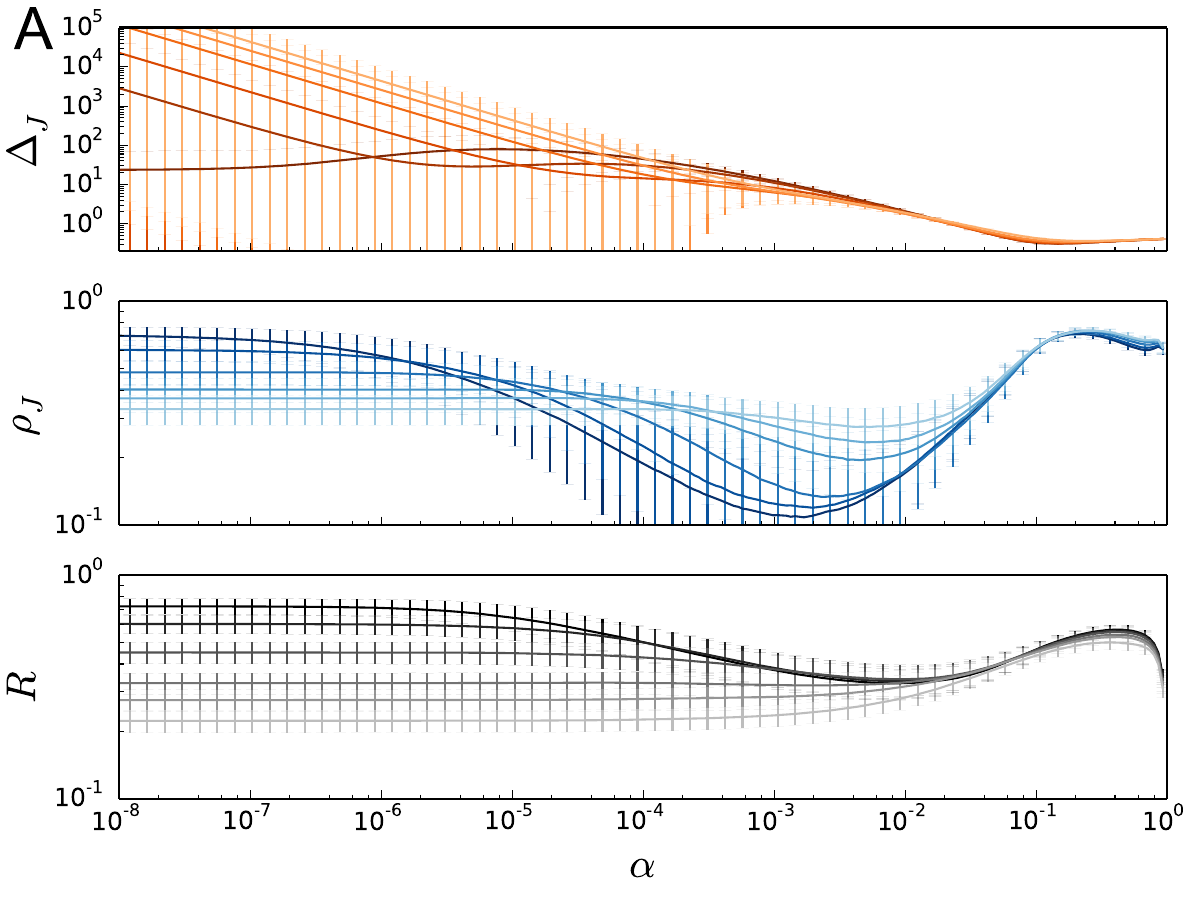}%{xl-er2-hmu0-jmu0-jsig3-Brange-pc-error}
\includegraphics*[width=3.5in]{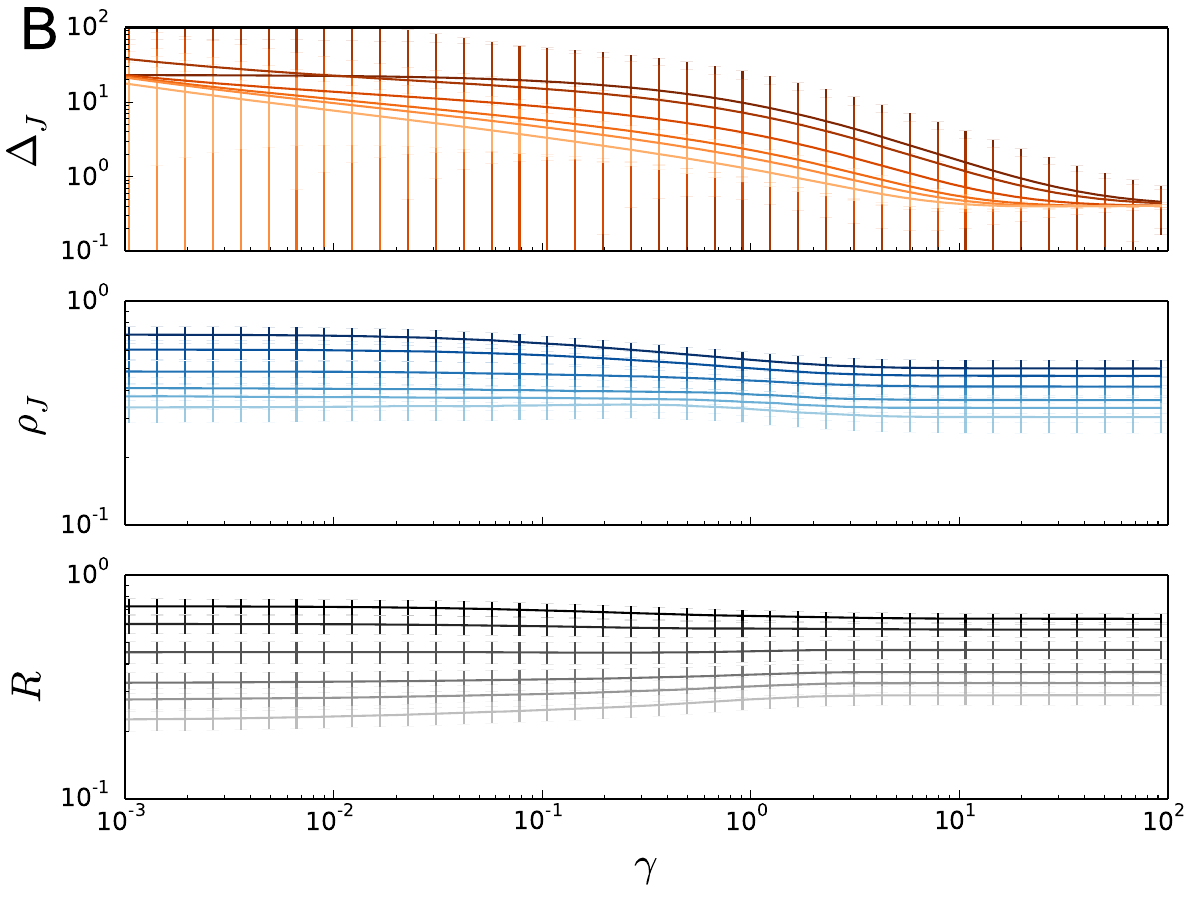}%{xl-er2-hmu0-jmu0-jsig3-Brange-l2-error}
\caption{
Optimal values of the regularization strength are only weakly affected by sampling depth, even when varied over the full range from $B=500$ (\textit{lightest}) to $B=10^6$ (\textit{darkest}).
Trajectory of the RMS error $\Delta_J$ \eqref{RMSJ} (\textit{top}), rank correlation $\rho_J$ \eqref{rhoJ} (\textit{middle}), and fraction of nonzero couplings recovered $R$ (\textit{bottom}) as the pseudo-count $\alpha$ (\textbf{A}) and $L_2$-norm regularization strength $\gamma$ (\textbf{B}) is varied, averaged over $10^3$ sets of random couplings. Dashed lines mark $\alpha=\alpha^{MF}=0.2$ (\textit{black, labeled}), and values of $\alpha^B=1/B$ (\textit{shaded}), roughly where $\rho_J$ and $R$ begin to plateau for the pseudo-count. Each random set of interactions has all fields set to zero. The coupling network is an Erd\H{o}s-R\'enyi graph where edges are kept with probability $p=2/N$. Nonzero couplings are normally distributed with mean zero and standard deviation $\sigma_J=3$. Bars denote one half standard deviation over the sample.
\label{Brange}}
\end{center}
\end{figure*}

Analysis presented in Section~\ref{model} suggests an optimal value for the regularization strength needed to correct for errors introduced by the MF approximation, which is independent of the amount of data. In contrast, in a Bayesian framework the regularization strength should scale as $\alpha^B \sim 1/B$ as the sampling depth is increased, where $B$ is the number of independent samples, as described in Section~\ref{schemes}. Our simulation results agree with the former picture: the optimal regularization strength $\alpha$ minimizing the RMS error between the true and inferred couplings is nearly independent of the sampling depth $B$, even when the latter is varied over four orders of magnitude. This is demonstrated in Fig.~\ref{Brange} for a system with underlying interactions given by an Erd\H{o}s-R\'enyi graph, but the result is completely general, holding for every model we have considered. 

\begin{figure*}
\begin{center}
\includegraphics*[width=3.5in]{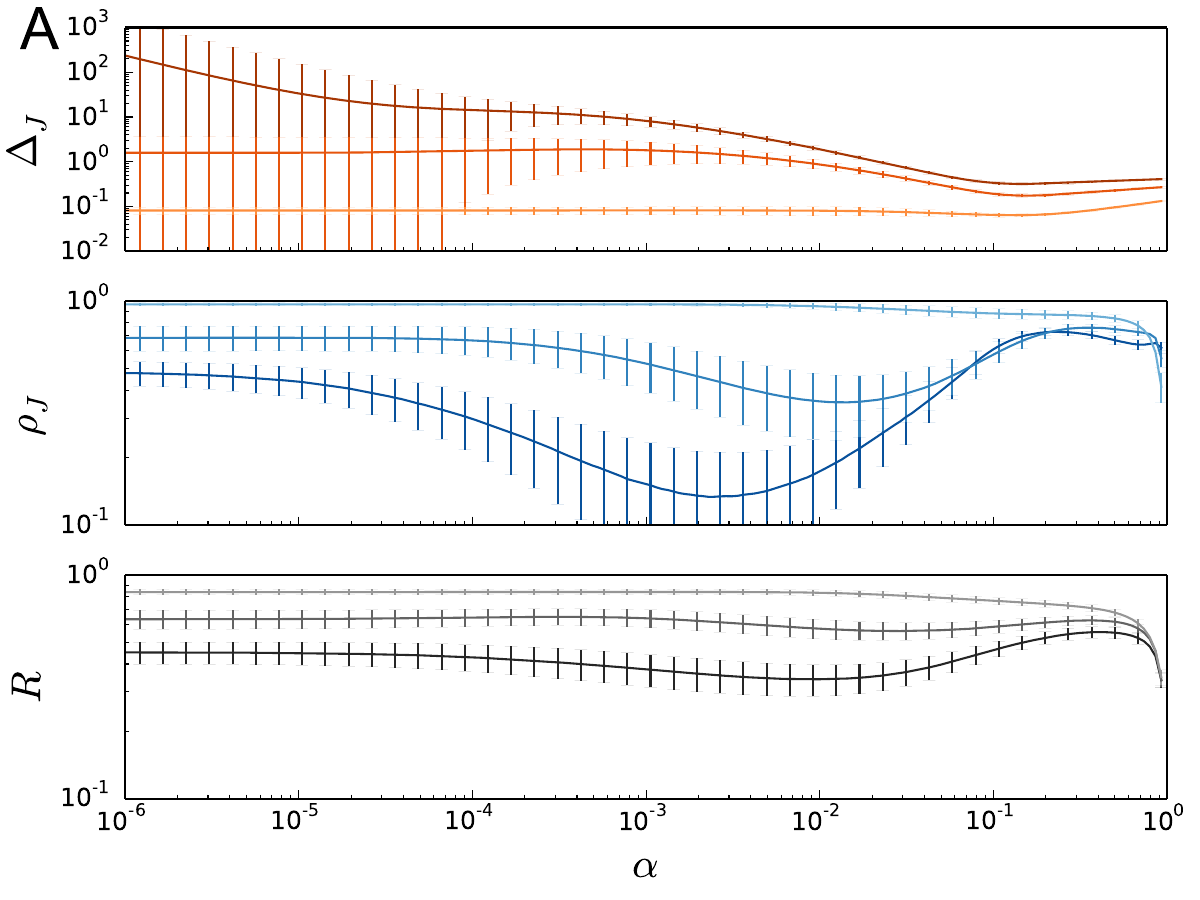}%{xl-er2-hmu0-jmu0-B10000-Jrange-pc-error}
\includegraphics*[width=3.5in]{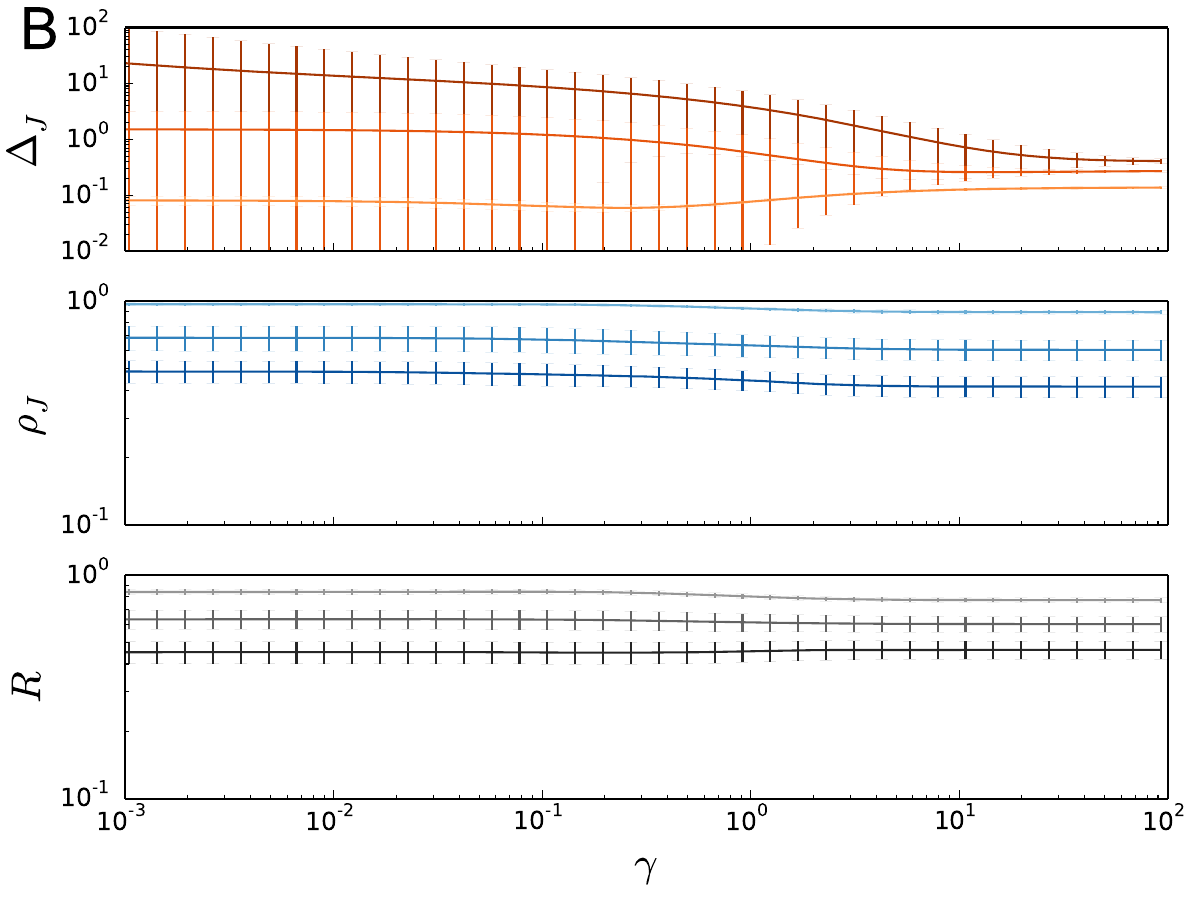}%{xl-er2-hmu0-jmu0-B10000-Jrange-l2-error}
\caption{
Varying the strength of the underlying interactions can shift the optimal value of the regularization strength.
Trajectory of the RMS error $\Delta_J$ \eqref{RMSJ} (\textit{top}), rank correlation $\rho_J$ \eqref{rhoJ} (\textit{middle}), and fraction of nonzero couplings recovered $R$ (\textit{bottom}) as the pseudo-count $\alpha$ (\textbf{A}) and $L_2$-norm regularization strength $\gamma$ (\textbf{B}) is varied, averaged over $10^3$ sets of random couplings, over a range of coupling distribution widths $\sigma_J=1$ (\textit{light}), $\sigma_J=2$ (\textit{medium}), and $\sigma_J=3$ (\textit{dark}). Dashed lines mark $\alpha=\alpha^{MF}=0.2$, and $\alpha^B=1/B=10^{-4}$, roughly where $\rho_J$ and $R$ plateau for the pseudo-count. Each random set of interactions has all fields set to zero. The coupling network is an Erd\H{o}s-R\'enyi graph where edges are kept with probability $p=2/N$. Nonzero couplings are normally distributed with mean zero and standard deviation $\sigma_J$. MF couplings were inferred from correlations computed from $B=10^4$ Monte Carlo samples of the true model. Bars denote one half standard deviation over the sample.
\label{Jrange}}
\end{center}
\end{figure*}

Independent of the value of $B$ we find that the value of the pseudo-count $\alpha$ which gives the best performance (smaller RMS error $\Delta_J$, largest rank correlation  $\rho_J$, and fraction $R$ of recovered nonzero couplings) is typically of the order of $\alpha^{MF}=0.2$, as computed in Section~\ref{sec:potts} for Ising spins. The rank correlation \eqref{rhoJ} (Fig.~\ref{Brange}A, middle) and the fraction of nonzero couplings recovered (Fig.~\ref{Brange}A, bottom) reach similar values with a small regularization strength $\simeq \alpha^{B}$ for very good sampling.
It is important to stress that for the pseudo-count the value of the optimal regularization strength is also largely independent of the strength of the interactions (Fig.~\ref{Jrange}). In the case of very weak interactions, MF inference is almost exact, and performance with or without pseudo-count is comparable.

\subsubsection{Comparison of pseudo-count and $L_2$ regularization performance on random graphs}

We now compare the performances of pseudocount and $L_2$ regularization.

\begin{figure*}
\begin{center}
\includegraphics*[width=.8\columnwidth]{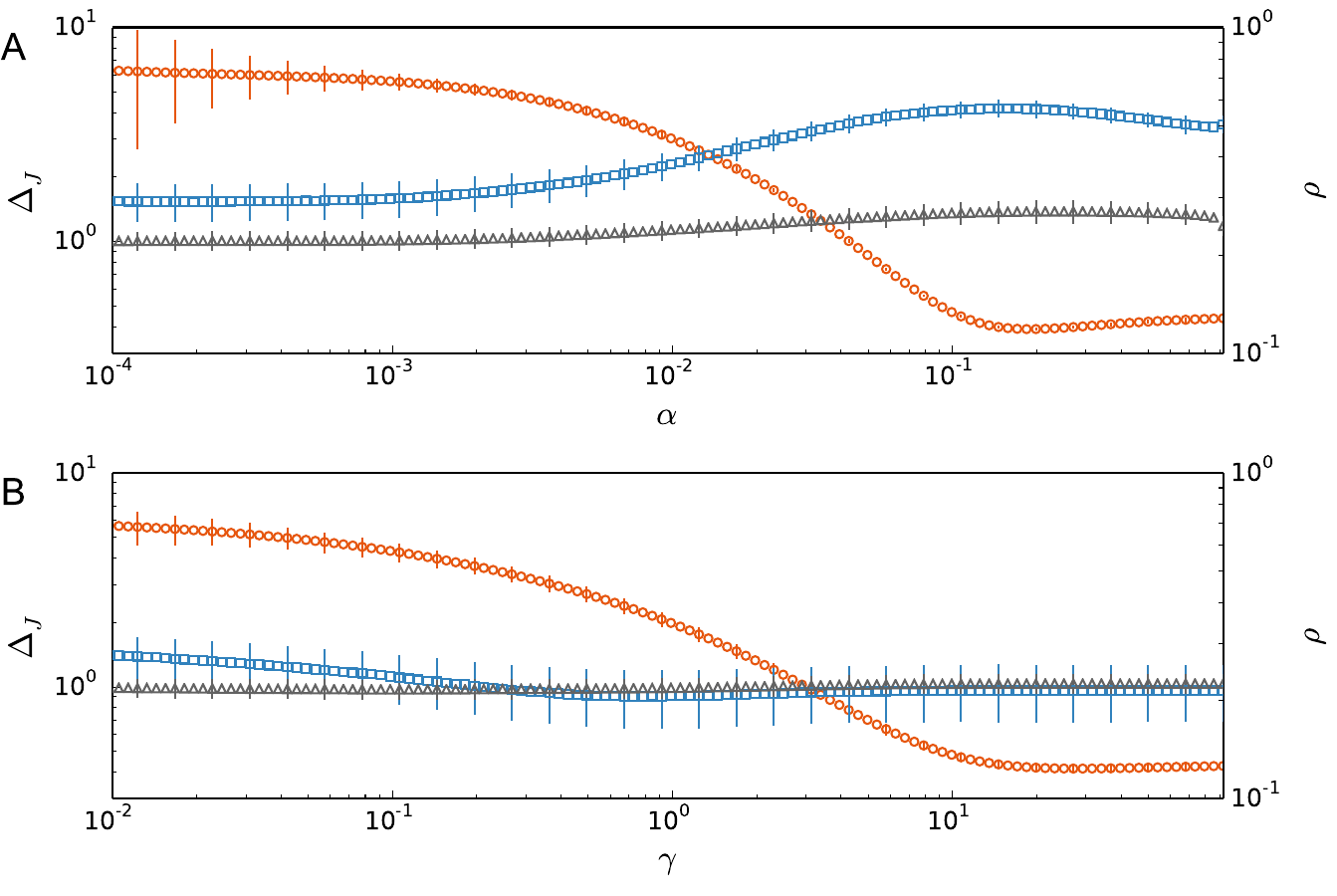}%{er4-hmu-5-jmu1-jsig2-B1000-error.pdf}
\caption{
Performance of the pseudo-count (\textbf{A}) and $L_2$-norm regularization (\textbf{B}) differ as a function of the regularization strength, particularly in the behavior of the rank correlation.
RMS error $\Delta_ J$ \eqref{RMSJ} (\textit{circles, left axis}), rank correlation $\rho_J$ \eqref{rhoJ} (\textit{squares, right axis}), and fraction of nonzero couplings recovered $R$ (\textit{triangles, right axis}), as a function of pseudo-count $\alpha$ and $L_2$-norm regularization strength $\gamma$, averaged over $10^3$ sets of random couplings. Dashed lines mark $\alpha=\alpha^{MF}=0.2$, and $\alpha^B=1/B=10^{-3}$, roughly where $\rho_J$ and $R$ plateau for the pseudo-count. Each random set of interactions has all fields $h=-5$. The coupling network is an Erd\H{o}s-R\'enyi graph where edges are kept with probability $p=4/N$. Nonzero couplings are normally distributed with mean $\bar{J}=1$ and standard deviation $\sigma_J=2$. MF couplings were inferred from correlations computed from $B=10^3$ Monte Carlo samples of the true model. Bars denote one half standard deviation over the sample.
\label{pc-l2-compare}}
\end{center}
\end{figure*}

Consistent with the analytical arguments presented in Section~\ref{model} for small systems, regularization improves the quality of couplings inferred via MF in large systems for a wide variety of underlying models. In particular, the RMS error $\Delta_J$ \eqref{RMSJ} is always improved by regularization unless both the couplings are weak and sampling is very good. The pseudo-count can substantially improve rank correlation $\rho_J$ \eqref{rhoJ} as well as the fraction of true nonzero couplings recovered $R$, particularly when sampling is poor. $L_2$-norm regularization has some effect on the rank correlation and fraction of nonzero couplings recovered, but tends to improve them only slightly compared to couplings inferred via MF with no regularization. In Fig.~\ref{pc-l2-compare} we compare the two methods for one example system on a relatively well sampled Erd\H{o}s-R\'enyi graph.
 Moreover in Fig.~\ref{Brange}B (middle and bottom panels ) we show that with the $L_2$ norm  the value of the rank correlation and the fraction of nonzero couplings depends, even at large regularization strengths, on the sampling depth. Moreover for poorly sampled systems a large regularization $\gamma$ does not improve as much as the one with the pseudo-count regularization.

Generally we find that the pseudo-count is well-suited to situations where the sampling depth is poor, and where the true interactions are strong. In such cases $\rho_J$ and $R$ can be achieve much larger values than with $L_2$ regularization, while maintaining similar RMS errors $\Delta_J$. This difference between the pseudo-count and $L_2$-norm regularization schemes can be understood through analysis of the $O(m)$ model, presented in Section~\ref{om}. Performance of the pseudo-count can be sensitive to changes in $\alpha$, but the optimal value of $\alpha$, while varying some with the strength of the true interactions, is generically of the same order as $\alpha^{MF}$ (Fig.~\ref{alphaq}). Additionally, $\Delta_J$ is typically small in the same range of $\alpha$ that maximizes the rank correlation and fraction of nonzero couplings recovered, making the pseudo-count particularly attractive in this regime.

$L_2$ regularization can offer modest advantages compared to the pseudo-count when sampling is very good, if the true couplings are not too strong. In these cases $L_2$ regularization can achieve slightly higher values of $\rho_J$ and $R$ at large values of $\gamma$, where the RMS error $\Delta_J$ is minimized. This method also has the advantage of being much less sensitive to the value of the regularization strength $\gamma$. 

\subsection{Results for the Potts model}\label{pottssimu}

In this section we consider the Potts model, with $q=5$ and $q=21$ symbols. We report below results for a one-dimensional interaction network, with $N=50$ sites; the qualitative conclusions we draw from the study of this model are in agreement with simulations on other interaction network geometries, not shown here. Two variants of this model will be studied, depending on how the Potts interactions vary between the symbols $\sigma_i,\sigma_{i+1}$ on neighboring sites along the chain:
\begin{itemize}
\item {\em Homogeneous variant:} For each pair of neighbors $i,i+1$, we draw randomly a number, $J_0$, uniformly between $-L$ and $L$, and set all the $q\times q$ couplings of the interaction matrix ${\bf J}_{i,i+1}$ through (\ref{potts7unb}). The process is repeated, independently, for all $N$ pairs of neighbors. The model is such that the $q$ Potts symbols have equal frequencies $p_i^a =\frac 1q$.

\item {\em Heterogeneous variants:} Extensions of the above model to non-equal frequencies can be easily obtained. To do so we consider a local field $h_i(a)$ on each site and symbol, which is also, for simplicity drawn uniformly at random from the $[-L,L]$ range. The explicit introduction of a field allows us to increase the bias between the frequencies of the $q$ symbols. This model will be called heterogenous-A model in the following.
We may, in addition to the introduction of random fields, draw randomly, for each pair of neighbors $i,i+1$, and for each pair of symbols $a,b$, a coupling $J_{i,i+1}(a,b)$, uniformly between $-L$ and $L$. Again, the process is repeated, independently, for all pairs of neighbors.We refer to this model as the heterogeneous-B model. 
 \end{itemize}
The values of the 1- and 2-point correlations are obtained from Monte Carlo simulations with the Hamiltonian (\ref{pottsH}) in the case of limited sampling (from $B$ configurations), and through a transfer matrix calculation in the case of perfect sampling ($B=\infty$) .
The following reports results of the MF inference,  with and without pseudo-count. In the latter case we use the strengths $\alpha=0.41$ for $q=5$ and $\alpha=0.75$ for $q=21$. Those choices correspond to the 'optimal' pseudo-count values found in Section \ref{sec:potts} (Fig.~\ref{alphaq}). 

We make sure that the coupling matrices ${\bf J}_{i,i+1}$ satisfy the zero-sum gauge: the sum of all couplings along each column and row of the coupling matrix vanish. The gauge is imposed through
\begin{equation}
J_{i,i+1}(a,b) \to J_{i,i+1}(a,b) - \frac{1}{q}\sum_{a=1}^{q} J_{i,i+1}(a,b) - \frac{1}{q}\sum_{b=1}^{q} J_{i,i+1}(a,b) + \frac{1}{q^2}\sum_{a,b=1}^{q}  J_{i,i+1}(a,b)\ .
\end{equation}
This choice allows us to compare the original and the inferred couplings.

\subsubsection{Importance of the bias in frequencies on the quality of inference}

\begin{figure}
\centering
\includegraphics[width=\textwidth]{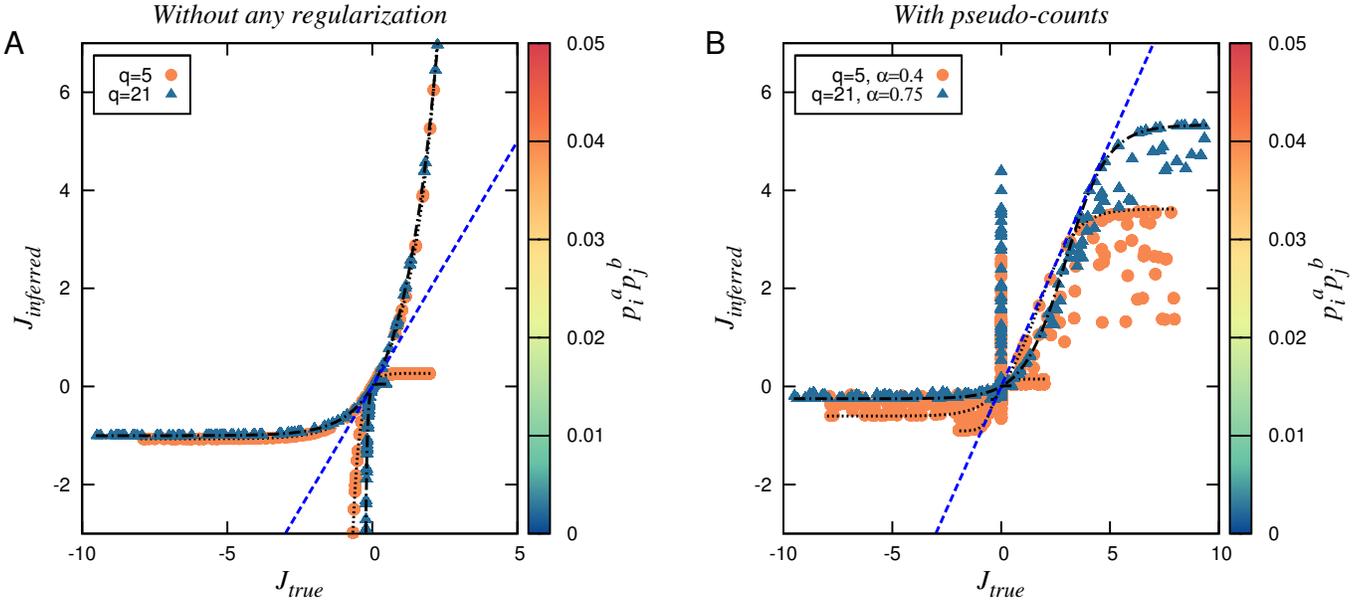}%{simupotts1-eps-converted-to.pdf}
\caption{Homogenous Potts model for $q=5$ (filled circles) and for $q=21$ (triangles) symbols; perfect sampling. {\bf (A)} No pseudo-count. {\bf (B)} With pseudo-count. Each panel shows results from three realizations with different sets of couplings  ($L=10$).  Black lines correspond to the analytical predictions of Section \ref{sec:potts}. Colors show values of $p_{i}^{a}p_{j}^{a}$, here equal to $q^{-2}$ for all interacting sites and for all symbols, see right scale. }  
\label{pottsFig1}
\end{figure}

In this section we consider the perfect sampling ($B=\infty$) case, in the presence or the absence of a pseudo-count regularization. We start with the homogeneous variant. The results of the MF inference are shown in Fig.~\ref{pottsFig1}A. We observe a perfect agreement with the analytical curves, see (\ref{potts7unb}) and (\ref{mfpotts}), if no regularization is present. The presence of a pseudo-count modifies the correlations between sites $i,j$. To understand the consequences of this modification, assume that true correlations decay exponentially with the distance between the sites. Consider three sites $i<j<k$ along the chain. Due to the exponential decay the  (connected) correlation $c_{ik}$ is equal to the product of the correlations between sites $c_{ij}$ and $c_{jk}$ (here for simplicity all Potts symbol indices are dropped). This equality expresses that the correlation is the result of interactions along the chain only. After the pseudo-count is introduced, all correlations are multiplied by a factor $(1-\alpha)$, and the equality $c_{ik}=c_{ij}\,c_{jk}$ no longer holds. The correlation $c_{ik}$ is now too large to be explained by a one-dimensional sequence of couplings from site $i$ to site $k$. As a result, many positive fictitious couplings are inferred to correct for this excess in correlation, corresponding to the points $(J=0,J^{PC}\ne 0)$ in the scatter plot of Fig.~\ref{pottsFig1}B. In turn, to correct for those extra fictitious couplings, the inferred values for the 'existing' (between adjacent sites) couplings are lowered with respect to their true values, see Fig.~\ref{pottsFig1}B. This effect could not be predicted from the two-spin calculation of Section \ref{sec:potts}, and is weaker if the true couplings are chosen from a smaller range, {\em i.e.} if $L$ is decreased.

\begin{figure}
\centering
\includegraphics[width=\textwidth]{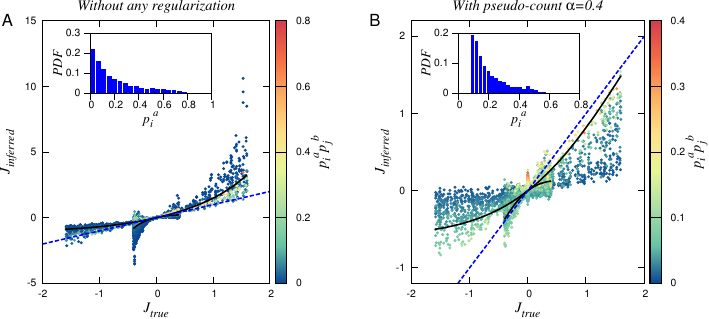}%{simupotts2-eps-converted-to.pdf}
\caption{Heterogenous-A Potts model for $q=5$ symbols, with perfect sampling. {\bf (A)} No pseudo-count. {\bf (B)} With pseudo-count. Each panel shows results from five realizations with different sets of couplings and fields ($L=2$). Insets: distributions of the frequencies $p_i^a$. Black lines correspond to the analytical predictions of Section \ref{sec:potts}. Colors show values of $p_{i}^{a}p_{j}^{a}$, see right scale.}  
\label{pottsFig2}
\end{figure}

\begin{figure}
\centering
\includegraphics[width=\textwidth]{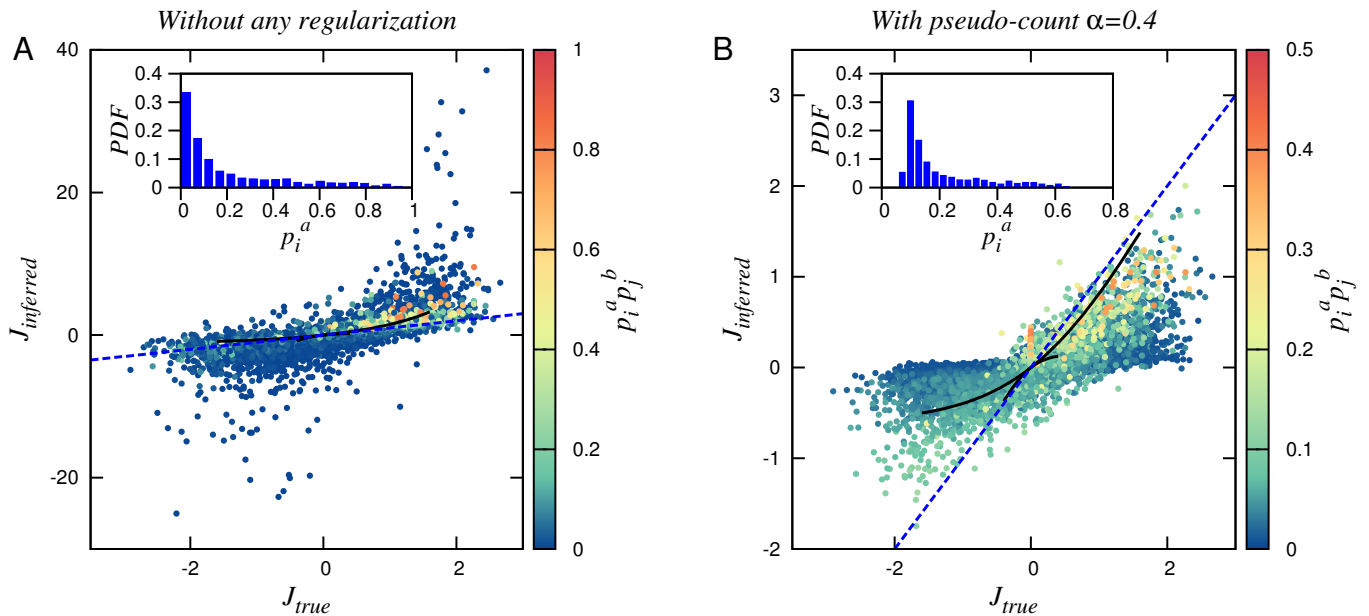}%simupotts4-eps-converted-to.pdf}
\caption{Heterogenous-B Potts model for $q=5$ symbols, with perfect sampling. {\bf (A)} No pseudo-count. {\bf (B)} With pseudo-count. Each panel shows results from five realizations with different sets of couplings and fields ($L=2$). Insets: distributions of the frequencies $p_i^a$. Black lines correspond to the analytical predictions of Section \ref{sec:potts}. Colors show values of $p_{i}^{a}p_{j}^{a}$, see right scale.}
\label{pottsFig4}
\end{figure}

Performance of MF inference for the heterogeneous variants  are shown in Figs.~\ref{pottsFig2}  and \ref{pottsFig4}. Comparison with the  toy-model analysis of Section \ref{sec:potts2}, see Fig.~\ref{figpotts9}, may be only qualitative here, because the heterogeneous models considered in the simulations have a richer distribution of frequencies $p_i^a$. In the toy model and in the presence of a pseudo-count, where frequencies can take only two values, four branches of couplings values appear in the regions $0<J^{PC}<J$ and  $J<J^{PC}<0$  of the  $(J,J^{PC})$ plane.  In the  heterogeneous models A and B the frequencies of the $q$ symbols are not bimodal, see inset of Figs.~\ref{pottsFig2} and \ref{pottsFig4}. We anticipate that branches will be less easy to identify, but will occupy the same regions of the plane in a dense way.

Results for the heterogeneous-A model are shown in Fig.~\ref{pottsFig2} for $q=5$ symbols; similar results were obtained for the $q=21$-symbol Potts model with and without the pseudo-count ($\alpha=0.75$). The agreement with the real couplings is generally better for couplings $J_{i,i+1}(a,b)$ corresponding to {\em conserved} sites and symbols, {\em i.e.} such that the product $p_{i}^{a}\,p_{i+1}^b$ is medium or large. This statement holds also in the presence of a pseudo-count, which improves the inference for pairs of sites and symbols with medium/large frequencies (green color in Fig.~\ref{pottsFig2}B). 
%This finding is in qualitative agreement with the toy-model analysis of Section \ref{sec:potts2}. 
The behavior of these inferred $J^{PC}$ couplings with a large or medium level of conservation is indeed
similar to the  branches $J^A$ $J^B$ from which we have tuned $\alpha_D$ (Fig.~\ref{alphaq}) in  the toy model for the homogeneous case.
Moreover, as seen in the homogeneous case (Fig.~\ref{pottsFig1}B), non-interacting but strongly conserved sites can generate fictitious and strong inferred couplings (red dots in Fig.~\ref{pottsFig2}B). One of the effects of the pseudo-count is,  indeed, to produce larger correlations between very conserved sites, and, in turn, nonzero inferred couplings between those sites.

Performances of the MF inference for the heterogeneous-B model are shown in Fig.~\ref{pottsFig4}. The distribution of  frequencies $p_i^a$ is more peaked at low values (inset of Fig.~\ref{pottsFig2}) than in the heterogeneous-A model (inset of Fig.~\ref{pottsFig4}).  The global picture is  similar to the one of Fig.~\ref{pottsFig2}, with an even wider dispersion.  Again, we find that couplings corresponding to medium or strongly conserved sites are generally better inferred than the ones corresponding to non-conserved sites for the mean field inference. In addition, in  the presence of a pseudo-count, nonzero couplings appear between non-adjacent and strongly conserved sites.

%As in the analytical calculation of Section \ref{sec:potts2} we observe that the inferred couplings in Fig.~\ref{pottsFig2}B and Fig.~\ref{pottsFig4}B occupy a large part of the $J^{PC}<J, J>0$ and $J^{PC}>J, J<0$ portions of the $(J,J^{PC})$ plane, compare with Fig.~\ref{figpotts9}, right panels. 

\subsubsection{Effects of finite sampling and reconstruction of the network structure}

\begin{figure}
\centering
\includegraphics[width=\textwidth]{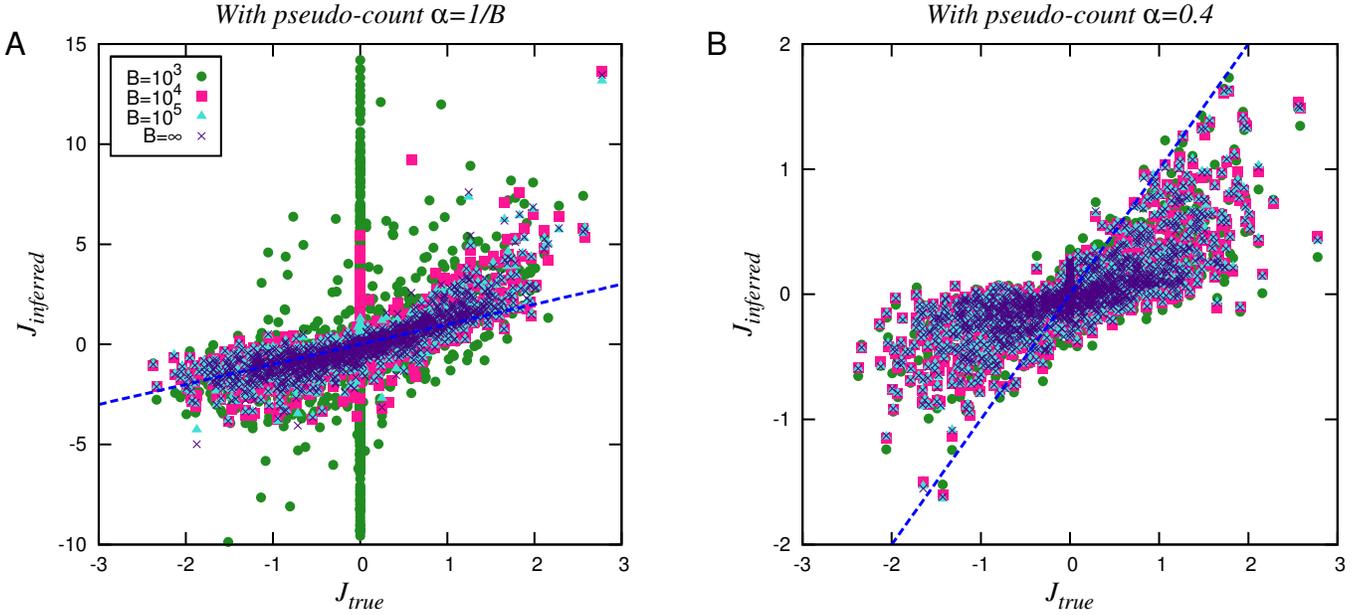}%simupotts5-eps-converted-to.pdf}
\caption{Heterogenous-B Potts model for $q=5$ symbols, for various depths of sampling. {\bf (A)} Small pseudo-count $\alpha=\alpha^B=\frac{1}{B}$. {\bf (B)} Large (optimal) pseudo-count $\alpha=\alpha^{MF}=0.4$. Each panel shows results from one realization of the Potts model with random couplings and fields ($L=2$), and three sets of $B$ sampled configurations (for finite $B$). 
%Insets: distributions of the frequencies $p_i^a$. 
%Black lines correspond to the analytical predictions of Section \ref{sec:potts}. 
%Colors show values of $p_{i}^{a}p_{j}^{a}$, see right scale.
}  
\label{pottsFig5}
\end{figure}

We now study the effect of finite sampling on MF inference for the heterogeneous-B model (Fig.~\ref{pottsFig5}). The errors on the inferred MF couplings are strongly affected by the sampling size for weak values of the pseudo-count, {\em e.g.} the large peak in $J^{MF}\ne 0, J^{true}=0$ corresponding to non-adjacent sites in Fig.~\ref{pottsFig5}A. Remarkably, for strong pseudo-count (optimal value defined in Section~\ref{sec:potts}), limited sampling has little effect on the inference error (Fig.~\ref{pottsFig5}B), which seems to be due primarily to the poor performance of MF inference in the presence of a wide distribution of the local frequencies $p_i^a$. 

\begin{figure}
\centering
\includegraphics[width=\textwidth]{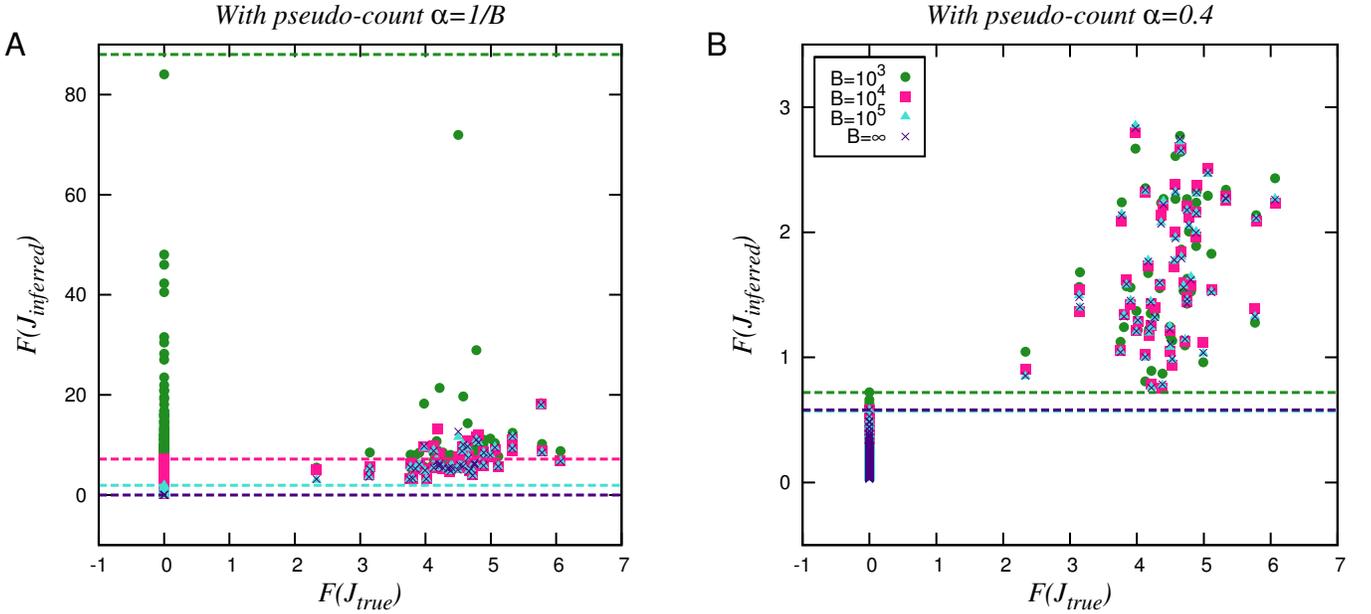}%simupotts6-eps-converted-to.pdf}
\caption{Scatter plot of the Frobenius norms of the inferred couplings vs. their true values for the pseudo-count strengths $\alpha=\frac 1B$ {\bf (A)} and $\alpha = 0.4$  {\bf (B)}. Same heterogeneous-B model and same conditions as in Fig.~\ref{pottsFig5}. Dotted lines locate the largest Frobenius norm corresponding to a pair of sites $(i,j)$ which are not neighbors on the one-dimensional graph, {\em i.e.} which have zero true coupling.}
\label{pottsFig6}  
\end{figure}

We present in Fig.~\ref{pottsFig6} the scatter plots of the Frobenius norm $(\sum_{a,b}J_{ij}(a,b)^2)^{1/2}$ for the same heterogeneous-B model as in Fig.~\ref{pottsFig5}. The Frobenius norm may serve as an estimator of the presence of a nonzero link in the interaction network. With a weak pseudo-count, $\alpha=\frac{1}{B}$, smaller sample sizes result in poorer performance. Many pairs $(i,j)$ with zero couplings give rise to Frobenius norms larger than the ones found for pairs of neighbors $(i,i+1)$, which have real nonzero couplings. This artifact implies that the graph structure cannot be correctly reconstructed, at least without having an estimation of the statistical error bars on single couplings due to the sampling noise \cite{Cocco:2012id}. However, with the optimal strength $\alpha=0.4$, finite sampling effects are better corrected for, and the correct structure of the graph is recovered. Even though the inferred couplings $J^{PC}_{ij}(a,b)$ differ from their true values $J_{ij}(a,b)$, the summation over the Potts symbols in the Frobenius norms seem to average out those errors, and to allow for a good inference of the underlying graph structure in the presence of a large pseudo-count. This result supports the use of large pseudo-counts in real applications such as protein contact predictions from covariation data \cite{Morcos:2011ct}.

\section{Analysis of the $O(m)$ model for large but finite $m$} \label{om}

In this Section we consider a large system of $N$ spins, and we want to estimate the error on the inferred couplings due to the MF approximation, and how this error can be corrected for with appropriate regularization. Estimating exactly this error would require that one solves exactly the inverse Ising or Potts model, which is computationally intractable. We therefore concentrate on an extension of the Ising model, the $O(m)$ model, which can be solved analytically for large $m$. The spin variables in the model are $m$-dimensional vectors, $\vec \sigma_i$, with squared norms constrained to be equal to $m$: $|\vec \sigma_i|^2=m, \forall i$. The case $m=1$ corresponds to Ising spins. We first recall how the properties of the $O(m)$ model can exactly solved in the infinite $m$ limit, and how a systematic expansion in powers of $1/m$ can be carried out. We then compute the error done by MF on the inverse $O(m)$ model and study to which extent these errors are compensated by pseudo-count and $L_2$ regularizations.

\subsection{Statistical mechanics of the inverse $O(m)$ model}

Given a set of interactions $J_{ij}$, the likelihood of a configuration $\{\vec \sigma_1,\vec \sigma_2, \ldots , \vec \sigma_N\}$ of the model is
\begin{equation}
p(\vec \sigma_1,\vec \sigma_2, \ldots , \vec \sigma_N) = \frac {\exp \big( \sum_{i<j} J_{ij}\, \vec \sigma_i \cdot \vec \sigma_j \big)}{Z(\{J_{ij}\})}\ ,
\end{equation}
where the partition function reads
\begin{equation}
Z(\{J_{ij}\}) =\int _{|\vec \sigma_1|^2=m} d\vec\sigma_1\ldots \int _{|\vec \sigma_N|^2=m} d\vec\sigma_N\; \exp \big( \sum_{i<j} J_{ij}\, \vec \sigma_i \cdot \vec \sigma_j \big) .
\end{equation}
In the above formulae $\cdot$ represents the dot product between two spin vectors. The correlation per spin component is defined through 
\begin{equation}\label{cij}
c_{ij} = \frac 1m \int _{|\vec \sigma_1|^2=m}d\vec\sigma_1\ldots \int  _{|\vec \sigma_N|^2=m}d\vec\sigma_N\;   p(\vec \sigma_1,\vec \sigma_2, \ldots , \vec \sigma_N)\, \vec \sigma_i \cdot \vec \sigma_j \ .
\end{equation}

To compute the partition function $Z$ we introduce imaginary-valued Lagrange multipliers $\lambda_i$ to enforce the constraint over the norm of $\vec \sigma _i$ for all $i=1,\ldots,N$. We obtain:
\begin{eqnarray}
Z(\{J_{ij}\}) &=&\int _{iR}\frac{d\lambda_1}{4\pi }\ldots \int _{iR}\frac{d\lambda_N}{4\pi }\int d\vec\sigma_1\ldots \int  d\vec\sigma_N\; \exp \bigg( \sum_{i<j} J_{ij}\, \vec \sigma_i \cdot \vec \sigma_j + \sum _j\frac {\lambda_j}2(m - \vec \sigma_j^2) \bigg) \nonumber \\
&=&2^{-N}(2\pi)^{N(m/2-1)} \int _{iR}d\lambda_1\ldots \int _{iR}d\lambda_N\exp \bigg( \frac m2 \sum_{j} \lambda_j -\frac m2 \log \det A(\lambda,J)  \bigg)
\end{eqnarray}
where $A(\lambda,J)$ is a $N\times N$ symmetric matrix, with diagonal elements $A_{ii}=\lambda_i$ and off-diagonal elements $A_{ij}=-J_{ij}$. For large $m$ we estimate the integral according to the saddle-point method: the values of the Lagrange multipliers $\lambda_i^*$ are such that the diagonal elements of the inverse matrix of $A$ are all equal to one: $[A(\lambda^*,J)^{-1}]_{ii}=1, \forall i$. Gaussian corrections to the saddle-point are easy to compute with the following expression for the log-likelihood of the data given the coupling matrix:
\begin{eqnarray}
L(J|c)&=& m\sum_{i<j} J_{ij} \,c_{ij} -\frac m2 \sum_i \lambda^*_i +\frac m2 \log \det A(\lambda^*,J) +\frac 1{2} \log\det HÊ(J)\ ,\nonumber \\
&=&  -\frac m2\, \hbox{\rm Trace}[A(\lambda^*,J)\; c] +\frac m2 \log \det A(\lambda^*,J) +\frac 1{2} \log\det HÊ(J)\ ,
\end{eqnarray}
where we have omitted an irrelevant $J$-independent additive constant, and the $N\times N$ symmetric matrix $H$ is defined through
\begin{equation}
H_{ij} (J)= \big( [A(\lambda^*,J)^{-1}]_{ij} \big)^2\ .
\end{equation}
$H$ is the point-wise square of a positive definite matrix; according to Schur product theorem it is itself a positive matrix.

We now consider the inverse $O(m)$ problem. We want to determine the coupling matrix $J=\{J_{ij}\}$ fulfilling the constraints  (\ref{cij}). To do so we maximize the log-likelihood $L$ with respect to $J$. In the infinite-$m$ limit the solution is simply $(J_\infty)_{ij}=-(c^{-1})_{ij}$, as expected for the Gaussian model, which is equivalent to the mean-field approximation. Deviations from the MF inference are found when $m$ is large but finite,
\begin{equation}\label{jijlargem}
\delta J_{ij}\equiv J_{ij} -(J_\infty)_{ij} = \frac 1m\, (H_\infty^{-1})_{ij}\, c_{ij} +O\left( \frac 1{m^2}\right)\ ,
\end{equation}
where 
\begin{equation}
(H_\infty) _{ij} =  ( c_{ij})^2\ .
\end{equation}
Expression (\ref{jijlargem}) is our ``exact'' value for the couplings given the correlation matrix. Below we study the accuracy of the MF prediction (in the presence of regularization) compared to this expression.

\subsection{Effect of regularization schemes}

We start with the $L_2$-regularization with link-dependent penalty, $\gamma_{ij}$, {\em i.e.}~we add a penalty term $-\frac 14 \sum_{i,j} \gamma_{ij} J_{ij}^2$ to the log-likelihood $L$ (recall that diagonal couplings $J_{ii}$  coincide here with $-\lambda_i^*$). We then extremize with respect to $J$, and ask for the change in $J$ resulting from the presence of this new $L_2$ penalty term to compensate exactly $\delta J$ given by (\ref{jijlargem}). A straightforward calculation leads to 
\begin{equation}\label{gm}
\gamma_{ij} =- \frac 1{m\,(c^{-1})_{ij}}  \sum_{k,\ell} c_{ik}\, (H_\infty^{-1})_{k\ell}\, c_{k\ell}\,  c_{\ell j}+O\left( \frac 1{m^2}\right) \ .
\end{equation}

We now repeat the approach with the pseudo-count regularization. We consider the general case of a link-dependent pseudo-count, of strength $\alpha_{ij}$. The off-diagonal entries $c_{ij}$ of the correlation matrix are now equal to $(1-\alpha_{ij})c_{ij}$, while the diagonal entries are unchanged: $c_{ii}=1$. In the $m\to\infty$ limit the change in the coupling $J_{ij}$ resulting from the presence of the pseudo-count is, to the first order in $\alpha$, 
\begin{equation}\label{gm2}
\delta J^{PC}_{ij} = - \sum _{k\ne\ell} (c^{-1})_{ik}\, \alpha_{k\ell}\, c_{k\ell}\, (c^{-1})_{\ell j} \ .
\end{equation}
We want to compute the values of the strengths $\alpha_{kl}$, with $k\ne l$, such that $\delta J^{PC}_{ij}$ and $\delta J_{ij}$ in (\ref{jijlargem}) sum up to zero for all $i\ne j$. The solutions are given by
\begin{equation}\label{am}
\alpha_{ij} =  \frac 1{m \; c_{ij}} \bigg[ \sum_{k,\ell} c_{ik} \, (H_\infty^{-1})_{k\ell}\, c_{k\ell}\, c_{\ell j}
- \sum _k c_{ik}\, d_k\, c_{kj}\bigg]+O\left( \frac 1{m^2}\right) \ ,
\end{equation}
where 
\begin{equation}
d_k= \sum _i (H_\infty^{-1})_{ki} \sum_{a,b} c_{ia} \, (H_\infty^{-1})_{ab} \, c_{ab}\, c_{bi}\ .
\end{equation}
The presence of the second term in the brackets in (\ref{am}) ensures that the pseudo-count vanishes on the diagonal, {\em i.e.}~$\alpha_{ii}=0$.

Two conclusions can be drawn from the previous calculations:
\begin{itemize}
\item We find that the optimal penalties, with $L_2$ regularization and pseudo-count, do not vanish in the pefect sampling limit considered here. The need for regularization can therefore not be due to poor sampling only. More precisely, the optimal penalties are of the order of $\frac 1m$ for large $m$. Loosely speaking, they are proportional to the deviation from the Gaussian model (recovered when $m\to\infty$, for which MF inference is exact).
\item We also understand from the formulas above why uniform pseudo-count is generally better than uniform $L_2$ penalty, as found in Section \ref{sec-uniform}. In (\ref{am}) the pseudo-count strength scales as the inverse of $c_{ij}$, which saturates to 1 for very strongly correlated spins. In contrast, in (\ref{gm}), the penalty scales as the inverse of $c^{-1}_{ij}$, that is, as $1/{J_{ij}}$. This quantity is not bounded from below when the coupling increases. Hence we expect a much wider range of values for the optimal $\gamma$ coefficients than for the optimal $\alpha$ coefficients. Uniform $L_2$ penalties are therefore far away from being optimal.
\end{itemize}

\section{Conclusion}

The present paper summarizes our efforts to understand the empirically observed necessity of large regularization terms in the mean-field inference of Ising or Potts interaction networks. In the usual Bayesian interpretation pseudo-count and $L_2$-norm regularization penalties are required in case of undersampling. As more data become available, the sampling noise becomes smaller, and so do the optimal values of the regularization terms. A combination of analytical and numerical evidences suggest that this interpretation is not correct for MF inference, and that the need for large regularization penalties rather comes from the non-Gaussian character of the variable statistics. In other words, large penalties, particularly for the pseudo-count, correct for the error in the inferred couplings introduced by MF. The importance of large regularization penalties to correct for errors introduced by the MF approximation is confirmed by analysis of the m-components $O(m)$ spin model: the optimal amplitude for the regularization scales, for large but finite $m$, as $\frac 1m$, which coincides with the measure of the discrepancy of the model with a Gaussian statistics.
  
In this work we also explored the performance of MF inference with different regularization schemes for a diverse set of underlying model systems.  In general cases both large pseudo-count and $L_2$-norm regularization can yield couplings which correlate well with the true couplings, the pseudo-count is easier to use (requires less finite-tuning) in general, especially when no knowledge of the true couplings exists to guide the choice of an appropriate regularization strength. Moreover a large pseudo-count gives extremely stable performance in the inference in case of limited sampling. For the Ising model case we find that the optimal pseudo-count to infer the network topology and the couplings values is  $\alpha^{MF}=0.2$,  independently of the sampling depth, of the values of the true interactions, and of the structure of the interaction network. This value corresponds to what analytically obtained for a toy model with only two spins.

The toy-model approach allows also to show that the mean-field inference of the coupling parameters $J_{ij}(a,b)$ in the Potts model case is poorer than in the Ising model case. The quality of the inference generally worsens with the heterogeneity among the frequencies of the $q$ Potts symbols (states). The introduction of a pseudo-count helps in reducing the errors on the inferred couplings $J_{ij}(a,b)$, especially for symbols $a$ and $b$ having medium/large frequencies on the site $i$ and $j$. For those couplings, an optimal value of the pseudo-count can be determined, which increases with $q$ and agrees to what is analytically found with the toy-model in the homogeneous case. Couplings attached to non-conserved sites are the ones with the largest inference errors, and cannot be reliably inferred even in the presence of a pseudo-count. Moreover the introduction of a pseudo-count may  lead to the prediction of nonzero couplings between non-interacting but very conserved sites, an artifact known in the context of the application of MF inference techniques to residue covariation in protein families \citep{Morcos:2011ct}. Even if for heterogeneous Potts models many couplings are poorly inferred, the introduction of a pseudo-count is helpful to improve the reconstruction of the interaction network structure in the case of limited sampling. This finding is, again, in good agreement with empirical results in the context of protein covariation.   

Our work could be extended in several ways. A potentially important finding of Section V is that there exist optimal value for the penalties, which depend on the empirical correlations, see (\ref{gm}) and (\ref{am}). It would be interesting to pursue this direction, and to see whether the introduction of link-dependent penalties could, indeed, improve the quality of MF inference in practical applications. Another issue of interest would be determine if and how the optimal regularization strengths depend on additional and specific constraints on the coupling matrix. A practical example is provided by the inverse Hopfield model \cite{Cocco:2011pr,Cocco:2013pl}, in which the rank $p$ of the interaction matrix $J$ is small compared to the system size. Last of all, it would be interesting to understand the observed stability of the inferred couplings against the sampling depth in the presence of a pseudo-count, see Fig.~\ref{pottsFig5} and Fig.~\ref{Jrange} for, respectively, the Potts and Ising cases. We expect performances to deteriorate when $B$ gets of the order of, or smaller than $N$ \cite{Cocco:2013pl}, but a more quantitative understanding of the minimal sampling depth required would be useful in practical applications.

\vskip .4cm
\noindent {\bf Acknowledgments.} We are grateful to M. Weigt for very useful discussions. We thank S. Rosay for her contribution to the analytical study of the $O(m)$ model in Section V. This work was partly funded by the Agence Nationale de la Recherche Coevstat project (ANR-13-BS04-0012-01).

\bibliographystyle{apsrev}
\bibliography{refs}

\end{document}